# Capacity Analysis for Continuous Alphabet Channels with Side Information, Part II: MIMO Channels

Majid Fozunbal, Steven W. McLaughlin, and Ronald W. Schafer*

School of Electrical and Computer Engineering
Georgia Institute of Technology
Atlanta, GA 30332-0250
{majid, swm, rws}@ece.gatech.edu

August 4, 2004

## Abstract

In this part, we consider the capacity analysis for wireless mobile systems with multiple antenna architectures. We apply the results of the first part to a commonly known baseband, discrete-time multiple antenna system where both the transmitter and receiver know the channel's statistical law. We analyze the capacity for additive white Gaussian noise (AWGN) channels, fading channels with full channel state information (CSI) at the receiver, fading channels with no CSI, and fading channels with partial CSI at the receiver. For each type of channels, we study the capacity value as well as issues such as the existence, uniqueness, and characterization of the capacity-achieving measures for different types of moment constraints. The results are applicable to both Rayleigh and Rician fading channels in the presence of arbitrary line-of-sight and correlation profiles.

---

*This work was supported in part by Texas Instruments Leadership University Program.

# 1   Introduction

Multiple antenna architectures have an increasingly important role to play in emerging wireless communication networks, particularly at base stations in cellular systems. Indeed, when used in conjunction with appropriately designed signal processing and coding algorithms, such architectures can dramatically enhance the performance of wireless mobile systems. However, in order to make efficient use of the resources, it is necessary to understand the fundamental limits of these systems.

Because of the time-varying nature of the channel in wireless mobile systems, the channel state (realization) is changing over time, called as *fading*, that imposes new challenges in determining the capacity of the channel. For this purpose, it is essential to know how much knowledge we have about the channel states either at the transmitter or at the receiver. In practice, depending on the application, we might have a range of scenarios from no channel state information (CSI) to full CSI. Hence, the capacity analysis and optimal coding strategies for different CSI scenarios could be quite different. For example, if full CSI is available at both the transmitter and the receiver, then the capacity-achieving input distribution is Gaussian, and the optimal encoder employs a power adaptation algorithm (water pouring) [1], [2], [3], [4]. In contrast, in the presence of full CSI at just the receiver, the capacity-achieving input distribution is Gaussian [5], [6], but the encoder uses the same average power over all time instances. This scenario is well investigated for multiple antenna channels in the presence of i.i.d. Rayleigh fading [7], [8], and recent interests in this area include determining the capacity and the capacity-achieving measures in the presence of arbitrary correlation and line-of-site fading components [9], [10], [11], [12], [13], [14].

Unlike these scenarios, the capacity and capacity-achieving distributions for fading channels in the absence of CSI, such as applications where the fading changes rapidly, are generally unknown even for the case of single-input single-output (SISO) channels. For multiple-input multiple-output (MIMO) channels with no CSI, Hochwald and Marzetta [15] have addressed the capacity problem for Rayleigh channels under certain assumptions on the SNR regime and on the ratio of the number of transmitters to the coherence time of the channels. For SISO channels, [16] was the first rigorous result in this area that addressed the characteri-zation of the capacity-achieving input distribution (subject to an average power constraint) for Rayleigh channels. Unlike Rayleigh channels, the capacity and capacity-achieving input distributions for Rician channels in the absence of CSI are barely touched. For low signal-to-noise ratio, [17] showed that the capacity-achieving input distribution (subject to second and fourth moment constraints) is discrete. Asymptotic upper bound and lower bound for the capacity of fading channels are derived in [18], subject to a maximum-power constraint. More results can be found in [19], and [20].

In this part, we use the results of Part I and study the capacity problem of MIMO channels in a unified manner, irrespective of the type of fading, the correlation profile, and the amount of available knowledge about the CSI at the receiver. More precisely, we study the capacity problem for AWGN channels, fading channels with full CSI at the receiver, fading channels with no CSI, and fading channels with partial CSI at the receiver. For each type of channels, we investigate its capacity as well as issues such as the existence, uniqueness, and characterization of the capacity-achieving measures of multiple antenna channels subject to different types of input moment constraints. The organization of this paper and a summary of our contributions are as follows.

In Section 2, we introduce the multiple antenna system setup. In Section 3, we address the capacity analysis for additive white Gaussian noise (AWGN) channels. Let $n$ and $m$ denote the number of transmit and receive antennas, respectively, and let $X = \mathbb{C}^n$ and $Y = \mathbb{C}^m$ denote the input and output alphabets of the channel. Suppose the channel realization is described by $\bar{H} \in \mathbb{C}^{m \times n}$. For moment constraints of type $\mathrm{E}\left(\|x\|_\eta^\eta\right)$ $(1 \leq \eta)$,[1] we show that capacity-achieving measure, $P_o$, exists uniquely. If $\eta > 2$, we show that the capacity-achieving measure has a bounded support with no interior point. In contrast, if $1 \leq \eta \leq 2$, then a necessary condition for $P_o$ is that for sufficiently large $y$ in the column space of $\bar{H}$, $\sup_{r \geq 0} e^{-r^2} P_o(\|y - \bar{H}x\|_2 \leq r\sigma) = O(e^{-\alpha \|y\|_\eta^\eta})$ for some $\alpha > 0$. For the case of $\eta = 2$, we also derive the capacity-achieving measure for these channels using Kuhn-Tucker conditions, where the result is the same as previously known results in [8].

In Section 4, we address the capacity analysis of MIMO fading channels with full CSI at the receiver for Rayleigh or Rician channels with arbitrary correlation profile. For moment constraints of type $\mathrm{E}\left(\|x\|_\eta^\eta\right)$ $(1 \leq \eta)$, we show that capacity-achieving measure, $P_o$, exists uniquely. If $\eta > 2$, we show that the capacity-achieving measure has a bounded support with no interior point. In contrast, if $1 \leq \eta \leq 2$, then a necessary condition for $P_o$ is that for almost every channel side information $v \in V = \mathbb{C}^{m \times n}$, $\sup_{r \geq 0} e^{-r^2} P_o(\|y - vx\|_2 \leq r\sigma) = O(e^{-\alpha(v)\|y\|_\eta^\eta})$ for sufficiently large $y$ in the column space of $v$ and for some positive function $\alpha : V \to \mathbb{R}^+$. For $\eta = 2$, we fully characterize the capacity-achieving measure for these channels, where our results reduces to the results of [8] for the case of isotropic Rayleigh channels.

---

[1]Refer to [21] for definition of $\eta$-norm.

In Section 5, we address the capacity analysis of MIMO fading channels with no CSI at the receiver for Rayleigh or Rician channels with arbitrary correlation profile. For moment constraints of type $\mathrm{E}\left(\|x\|_\eta^\eta\right)$ $(1 \leq \eta)$, we show that capacity-achieving measure, $P_o$, exists uniquely. If $\eta > 2$, we show that the capacity-achieving measure has a bounded support with no interior point. In contrast, if $1 \leq \eta \leq 2$, a necessary condition for the capacity-achieving measure is

$$P_o(\|y\|_2 \leq \|x\|_2, \mathcal{R}(y) \subseteq \mathcal{R}(x)) = O(e^{-\alpha\|y\|_\eta^\eta}),$$

where $\mathcal{R}(\cdot)$ denote the row space of a matrix, and $\alpha > 0$ is a constant.

In Section 6, we address the capacity analysis of MIMO fading channels with partial CSI at the receiver. We consider a certain class of estimators where the channel side information is jointly Gaussian with the channel realization. For moment constraints of type $\mathrm{E}\left(\|x\|_\eta^\eta\right)$ $(1 \leq \eta)$, we show that capacity-achieving measure, $P_o$, exists uniquely. If $\eta > 2$, we show that the capacity-achieving measure has a bounded support with no interior point. In contrast, if $1 \leq \eta \leq 2$, a necessary condition for the capacity-achieving measure is

$$\forall v \in V, \ P_o(\|y\|_2 \leq \lambda_{\min}(v)\|x\|_2, \mathcal{R}(y) \subseteq \mathcal{R}(x)) = O(e^{-\alpha(v)\|y\|_\eta^\eta}),$$

where $\mathcal{R}(\cdot)$ denotes the row space of a matrix, $V$ denotes the state information space, $\lambda_{\min}(v)$ denotes the minimum eigenvalue of the covariance of channel realization conditioned on $v$, and $\alpha : V \to \mathbb{R}^+$ is a positive function.

Finally, Section 7 states some concluding remarks along with some directions for future research.

## 2 General System Model

We assume a wireless communication system employing $n$ transmit and $m$ receive antennas, where the baseband model of the channel is described by a discrete-time model as follows. For each pair of transmit and receive antennas, $(s, r)$, the path from transmit antenna $s$ to receive antenna $r$ is represented by a complex symbol $h_{rs}$ called the *path gain*. Let $H \in \mathbb{C}^{m \times n}$ denote an $m \times n$ matrix with $h_{rs}$'s as its entries which is known as the *channel state or channel realization*.[2] Correspondingly, we denote the space of all possible states as $\boldsymbol{H} = \mathbb{C}^{m \times n}$.

We assume a block channel model where in each block the channel is used $L \geq 1$ times which is called the *block-length*. In each channel use, all antennas are used simultaneously, and $n$ complex symbols are transmitted through $n$ transmit antennas. We assume the channel is governed by a linear statistical model as follows.

Let $X = \mathbb{C}^{n \times L}$ and $Y = \mathbb{C}^{m \times L}$ denote the input and output alphabets, respectively. At each block $k$, a matrix $x_k \in X$ is transmitted through the transmit antennas and a matrix $y_k \in Y$ is received in receiver which is described by

$$y_k = H_k x_k + z_k, \tag{1}$$

where $z_k \in \mathbb{C}^{m \times L}$ denotes the additive noise and at block $k$. We assume that the noise matrices are temporally independent, with identically distributed (i.i.d.) complex normal entries which have zero mean and variance $\sigma^2$, i.e., $\mathcal{CN}(0, \sigma^2)$. We assume that the channel state remains unchanged during each block, but it might change after each block. We assume that at each block, there exists an element $v_k$ available at the receiver that gives some information about the channel state $H_k$. This enables us to consider a broad range of channel state information (CSI) scenarios from no CSI where $v_k$ and $H_k$ are statistically independent, to full CSI where conditioned on $v_k$ there is no uncertainty about $H_k$. We assume that $v_k$ belongs to a Borel-measurable space $V$ and there exists a joint measure $Q \circ R$ on $\boldsymbol{H} \times V$ and the input alphabet $X$ is statistically independent from $\boldsymbol{H} \times V$. Note that to fully characterize the statistical properties of the channel, we need to specify the joint probability measure $Q \circ R$, however, we postpone this to later sections where we address it in different scenarios.

---

[2]To clarify any confusion which might be raised by this notion in comparison to our notion in Part I, one should note that channel realization matrix, $H$, denotes the state of the channel which was previously shown as $s$. The reason of this change is to comply with the common notion in the literature of MIMO channels.

We assume that there exists a nonnegative continuous function $g : X \to \mathbb{R}^+$ and a positive value $\Gamma > 0$, such that for a $K$-length block code, the codewords are chosen to satisfy $\frac{1}{K} \sum_{k=1}^{K} g(x_k) \leq \Gamma$. Motivated by practical scenarios, we consider $g(x) = \|x\|_\eta^\eta$ $(1 \leq \eta < \infty)$, where the most common choice is $g(x) = \|x\|_2^2$ (the Frobenius norm) and $\Gamma$ is the average energy per block. By the Law of Large Numbers [22, p. 325], as $K$ grows to infinity, this is equivalent to assuming that the empirical measures of codes are obtained from a set of input probability measures which are characterized by a continuous positive function $g(x)$ together with a real value $\Gamma > 0$ as follows,

$$\mathscr{P}_{g,\Gamma}(X) = \left\{ P \in \mathscr{P}(X) \mid \int g(x) dP \leq \Gamma \right\}.$$

Note that for the choice $g(\cdot) = \|x\|_\eta^\eta$ $(1 \leq \eta < \infty)$,[3] one can easily verify that $g$ satisfy the hypothesis of Lemma 3.1 of Part I, hence, $\mathscr{P}_{g,\Gamma}(X)$ is weak* compact.

It just remains to specify the generic statistical law that governs the channel, that is to describe $W(\cdot|x, H)$. For this purpose, recall that the additive white noise is a complex normal random matrix with i.i.d. components. Hence, the channel is described by the conditional measure, $W(\cdot|x, H)$, which is absolutely continuous with respect to the Lebesgue measure, i.e., $W(\cdot|x, H) \ll \mu_Y$, with the density function

$$f(y|x, H) = \frac{1}{\pi^{mL} \sigma^{2mL}} e^{-\frac{\|y - Hx\|_2^2}{\sigma^2}}. \tag{2}$$

Let define an auxiliary measure $T$ as follows,

$$\forall E \in \mathcal{B}_Y, \quad T(E) = \frac{1}{\beta \pi^{mL} \sigma^{2mL}} \int_E e^{-\frac{\alpha^2 \|y\|_2^p}{\sigma^2}} d\mu_Y, \tag{3}$$

where $1 < p \leq 2$ and $0 < \alpha < 1$ are free variables and $\beta > 0$ is chosen such that $T$ has a unit norm. Since (2) is nonzero, one can verify that $W(\cdot|x, H) \ll T$ for all $x$ and $H$ and the density function of $W(\cdot|x, H)$ with respect to $T$ is described by

$$f_T(y|x, H) \triangleq \frac{dW(\cdot|x, H)}{dT} = \beta e^{\frac{\alpha^2 \|y\|_2^p - \|y - Hx\|_2^2}{\sigma^2}}. \tag{4}$$

---

[3]Refer to [21] for the definition of $\eta$-norm.

# 3 Additive White Gaussian Noise Channels

In this section, we consider a class of wireless channels where the physical medium between the transmitter and receiver remains unchanged throughout the communication. We assume that the channel is governed by a linear model as (1) and the channel state (realization) is $H_k = \bar{H}$ for all blocks, where $\bar{H}$ is known both at the transmitter and the receiver. Thus, we characterize the channel just by the matrix $\bar{H}$. To emphasize that a channel is Gaussian, we use a subscript "G" and denote the channel by $W_G(\bar{H})$. Since $\bar{H}$ is known, there is no advantage of taking $L > 1$. Hence, throughout this section, we assume that $L = 1$.

In the framework of the general system model, we consider that $V = \boldsymbol{H}$ where the probability measure $Q \circ R$ on $\boldsymbol{H} \times V$ is a point mass measure (Dirac measure) [21] at $(\bar{H}, \bar{H})$ that can be explained as follows. Since there is no uncertainty on the channel realization with the knowledge of $v$ at the receiver, the conditional probability measure $Q_v$ on $\boldsymbol{H}$ is a point mass measure at $H = v$. Moreover, because the channel state remains unchanged, the measure $R$ on $V$ is a point mass measure at $v = \bar{H}$. As a result, we observe that the channel can be simply described by measure $W_{Q_{\bar{H}}}(\cdot|x) \ll T$ with the density function

$$f_{T,Q_{\bar{H}}}(y|x) = \int f_T(y|x,H) dQ_{\bar{H}} = \beta e^{\frac{\alpha^2 \|y\|_2^p - \|y - \bar{H}x\|_2^2}{\sigma^2}}. \tag{5}$$

For every $P \in \mathscr{P}_{g,\Gamma}(X)$, it can be verified that $PW_{Q_{\bar{H}}} \ll T$ with the density function

$$f_{T,P,Q_{\bar{H}}}(y) = \beta \int e^{\frac{\alpha^2 \|y\|_2^p - \|y - \bar{H}x\|_2^2}{\sigma^2}} dP. \tag{6}$$

As a result, we simplify the expression of the mutual information (6 of Part I) for the additive white Gaussian channels as

$$I(P, W_G(\bar{H})) = \iint f_{T,Q_{\bar{H}}}(y|x) \log_2 f_{T,Q_{\bar{H}}}(y|x) dT dP - \int f_{T,P,Q_{\bar{H}}}(y) \log_2 f_{T,P,Q_{\bar{H}}}(y) dT. \tag{7}$$

Note that both of the terms on the right-hand side (RHS) in (7) are finite.

## 3.1 Properties of mutual information

In this subsection, we address some analytical properties of the mutual information function of the Gaussian channels. This includes properties such as strict concavity and continuity of the mutual information, which are used in the capacity analysis of channels.

Recall that Proposition 3.3 of Part I addresses the strict concavity of the mutual infor-mation in general. Since the strictness property is essential to addressing the uniqueness of the capacity-achieving probability measure, simpler arguments are of interest. Using the following observation, we address this issue.

**Observation 3.1.** *For every $P \in \mathscr{P}_{g,\Gamma}(X)$, $f_{T,P,Q_{\bar{H}}}(y)$ is continuous and nonzero over $Y$.*

*Proof.* By (5), the continuity and positiveness of $f_{T,Q_{\bar{H}}}(y|x)$ are obvious. The positiveness of $f_{T,Q_{\bar{H}}}(y|x)$ implies the positiveness of $f_{T,P,Q_{\bar{H}}}(y)$. To prove the continuity, suppose that $y_n \to y$ in the Euclidean norm. Then,

$$\begin{aligned}
\lim_n f_{T,P,Q_{\bar{H}}}(y_n) &= \beta \lim_n \int e^{\frac{\alpha^2 \|y_n\|_2^p - \|y_n - \bar{H}x\|_2^2}{\sigma^2}} dP \\
&= \beta \lim_n e^{\frac{\alpha^2 \|y_n\|_2^p}{\sigma^2}} \lim_n \int e^{-\frac{\|y_n - \bar{H}x\|_2^2}{\sigma^2}} dP \\
\text{By DCT } &= \beta e^{\frac{\alpha^2 \|y\|_2^p}{\sigma^2}} \int e^{-\frac{\|y - \bar{H}x\|_2^2}{\sigma^2}} dP \\
&= f_{T,P,Q_{\bar{H}}}(y).
\end{aligned}$$

This proves the continuity of $f_{T,P,Q_{\bar{H}}}(y)$ over $Y$. $\square$

We define that two probability measures $P_1, P_2 \in \mathscr{P}_{g,\Gamma}(X)$ are *equivalent* over $W_G(\bar{H})$, if they induce the same output probability measure. Equivalently, two input measures are equivalent over $W_G(\bar{H})$ if $f_{T,P_1,Q_{\bar{H}}}(y) = f_{T,P_2,Q_{\bar{H}}}(y)$ for all $y \in Y$.

**Proposition 3.1 (Strict concavity).** *The mutual information of a Gaussian channel $W_G(\bar{H})$ is strictly concave with respect to the convex combination of two input measures $P_1$ and $P_2$, unless they are equivalent over $W_G(\bar{H})$.*

*Proof.* By Observation 3.1, if there exists $y \in Y$ such that $f_{T,P_1,Q_{\bar{H}}}(y) \neq f_{T,P_2,Q_{\bar{H}}}(y)$, then there exists a neighborhood $U_y \subset Y$ of $y$ with that property. Then, by definition of $T$ (3), $T(U_y) > 0$. This means that the set $U_y \times \{\bar{H}\}$ complies the requirements of Proposition 3.3 of Part I. As a result, the mutual information function of a Gaussian channel is strictly concave with respect to the convex combination of $P_1$ and $P_2$ if and only if there exists $y \in Y$ such that $f_{T,P_1,Q_{\bar{H}}}(y) \neq f_{T,P_2,Q_{\bar{H}}}(y)$. By definition of equivalency for input probability measures, we deduce the assertion. $\square$

Another important property of the mutual information in capacity analysis is its weak* continuity, which is stated as follows.

**Proposition 3.2 (Continuity).** *The mutual information of any Gaussian channel $W_G(\bar{H})$ is weak\* continuous over $\mathscr{P}_{g,\Gamma}(X)$.*

*Proof.* It suffices to check if the hypothesis of Theorem 3.3 of Part I is satisfied. We prove the theorem for $g(x) = \|x\|_2^2$, and the proof for other choices of $g$ is also similar. To prove hypothesis (a) of Theorem 3.3 of Part I, we proceed as follows. Assume that $\alpha < 1$ and $p < 2$, which are fixed, and let

$$A_c = \{(x,y) \in X \times Y | f_{T,Q_{\bar{H}}}(y|x) \left|\log_2 f_{T,Q_{\bar{H}}}(y|x)\right| > c^2\}.$$

Let define

$$B_c = \{(x,y) \in X \times Y | f_{T,Q_{\bar{H}}}(y|x) > c\}.$$

For sufficiently large $c > 0$, e.g. $c > 10$, it can be observed that $A_c \subset B_c$. But the condition $f_{T,Q_{\bar{H}}}(y|x) > c$ is equivalent to the condition

$$\alpha^2 \|y\|_2^p - \|y - \bar{H}x\|_2^2 > \sigma^2 \ln \frac{c}{\beta}.$$

Hence, if we define

$$E_c = \{(x,y) \in X \times Y | \alpha^2 \|y\|_2^p > \sigma^2 \ln \frac{c}{\beta}\},$$

we can deduce that for sufficiently large $c > 0$, we would have $A_c \subseteq B_c \subseteq E_c$. Let $\omega(c) = \left(\frac{\sigma^2}{\alpha^2} \ln \frac{c}{\beta}\right)^{1/p}$. As a result, for every $P \in \mathscr{P}_{g,\Gamma}(X)$

$$\iint_{A_c} f_{T,Q_{\bar{H}}}(y|x) \left|\log_2 f_{T,Q_{\bar{H}}}(y|x)\right| dTdP \leq \iint_{E_c} f_{T,Q_{\bar{H}}}(y|x) \left|\log_2 f_{T,Q_{\bar{H}}}(y|x)\right| dTdP$$

$$\leq |\log_2 \beta| \iint_{E_c} \frac{1}{\pi^m \sigma^{2m}} e^{-\frac{\|y-\bar{H}x\|_2^2}{\sigma^2}} d\mu_Y dP$$

$$+ \frac{\log_2 e}{\sigma^2} \iint_{E_c} \frac{\alpha^2 \|y\|_2^p}{\pi^m \sigma^{2m}} e^{-\frac{\|y-\bar{H}x\|_2^2}{\sigma^2}} d\mu_Y dP$$

$$\leq \left(\frac{|\log_2 \beta|}{\omega^2(c)} + \frac{\log_2 e}{\sigma^2 \omega^{2-p}(c)}\right) \iint \frac{\|y\|_2^2}{\pi^m \sigma^{2m}} e^{-\frac{\|y-\bar{H}x\|_2^2}{\sigma^2}} d\mu_Y dP$$

$$= \left(\frac{|\log_2 \beta|}{\omega^2(c)} + \frac{\log_2 e}{\sigma^2 \omega^{2-p}(c)}\right) \left(m\sigma^2 + \int \|\bar{H}x\|_2^2 dP\right)$$

$$\leq \left(\frac{|\log_2 \beta|}{\omega^2(c)} + \frac{\log_2 e}{\sigma^2 \omega^{2-p}(c)}\right) (m\sigma^2 + \|\bar{H}\|^2 \Gamma)$$

Since $\omega(c) \to \infty$ as $c \to \infty$, we can deduce that

$$\lim_{c \to \infty} \sup_{P \in \mathscr{P}_{g,\Gamma}(X)} \int_{A_c} f_{T,Q_{\bar{H}}}(y|x) \left|\log_2 f_{T,Q_{\bar{H}}}(y|x)\right| dTdP = 0.$$

This implies that the hypothesis (a) of Theorem 3.3 of Part I holds. To verify the hypothesis (b), for a fixed $y$, let $D_c = \{x \in X : f_{T,Q_{\bar{H}}}(y|x) > c\}$. Then,

$$\sup_{P \in \mathscr{P}_{g,\Gamma}(X)} \int_{D_c} f_{T,Q_{\bar{H}}}(y|x) dP \leq \frac{1}{c} \sup_{P \in \mathscr{P}_{g,\Gamma}(X)} \int (f_{T,Q_{\bar{H}}}(y|x))^2 dP$$

$$= \frac{\beta^2}{c} \sup_{P \in \mathscr{P}_{g,\Gamma}(X)} \int e^{2\frac{\alpha^2 \|y\|_2^p - \|y - \bar{H}x\|_2^2}{\sigma^2}} dP$$

$$\leq \frac{\beta^2}{c} e^{2\frac{\alpha^2 \|y\|_2^p}{\sigma^2}} \to 0, \text{ as } c \to \infty.$$

Thus, since both hypotheses of Theorem 3.3 of Part I hold, the mutual information is weak* continuous. □

## 3.2 Capacity analysis

In this subsection, we address issues on the existence, the uniqueness, and the characterization of the capacity-achieving measures for Gaussian channels.

**Lemma 3.1 (Existence).** *For any Gaussian channel $W_G(\bar{H})$ and the set $\mathscr{P}_{g,\Gamma}(X)$, there exists a measure $P_o \in \mathscr{P}_{g,\Gamma}(X)$ that achieves the capacity of the channel $W_G(\bar{H})$.*

*Proof.* Proposition 3.2 states the weak* continuity of the mutual information over $\mathscr{P}_{g,\Gamma}(X)$. Since $\mathscr{P}_{g,\Gamma}(X)$ is weak* compact, by Proposition 4.1 of Part I, we conclude the assertion. □

Lemma 3.1 states the existence of the capacity-achieving measure over Gaussian channels. To address its uniqueness, we use an earlier result on strict concavity of mutual information, i.e., Proposition 3.1.

**Lemma 3.2 (Uniqueness).** *For any Gaussian channel $W_G(\bar{H})$ and the set $\mathscr{P}_{g,\Gamma}(X)$, the capacity-achieving measure is unique up to the equivalency of input measures.*

*Proof.* Suppose there exist distinct probability measures $P_o$ and $P_*$ that achieve the capacity. By Proposition 3.1, the mutual information is strictly concave with respect to their convex combination. This means that any measure in the form $\alpha P_o + (1-\alpha) P_*$ achieves a higher mutual information that is a contradiction to optimality of $P_o$ and $P_*$. Thus, $P_o$ and $P_*$ must be equivalent. □

So far, in this subsection, we have shown the existence and the uniqueness of the capacity-achieving measure over AWGN channels. It remains to provide some insight to the characterization of such input measure.

**Proposition 3.3.** *Let $g(\cdot) = \|\cdot\|_\eta^\eta$ for $1 \leq \eta$. If $\eta > 2$, the capacity-achieving measure has a bounded support with no interior point. In contrast, if $1 \leq \eta \leq 2$, then a necessary condition for $P_o$ is that for sufficiently large $y$ in the column space of $\bar{H}$, $\sup_{r \geq 0} e^{-r^2} P_o(\|y - \bar{H}x\|_2 \leq r\sigma) = O(e^{-\alpha \|y\|_\eta^\eta})$ for some $\alpha > 0$.*

*Proof.* Suppose $P_o W_{Q_{\bar{H}}}$ is the optimal capacity-achieving output measure. Applying Kuhn-Tucker condition, Theorem 4.3 of Part I, to our problem, we need to find a positive value $\gamma > 0$ such that

$$D(W_{Q_{\bar{H}}}(\cdot|x)\|P_o W_{Q_{\bar{H}}}) - \gamma \|x\|_\eta^\eta \leq C - \gamma \Gamma.$$

Using some straightforward mathematical manipulation, this results to

$$-m \log_2(\pi \sigma^2 e) - \int \frac{1}{(\pi \sigma^2)^m} e^{-\frac{\|y - \bar{H}x\|_2^2}{\sigma^2}} \log_2 f_{P_o, Q_{\bar{H}}}(y) d\mu_Y - \gamma \|x\|_\eta^\eta \leq C - \gamma \Gamma. \qquad (8)$$

The problem is now finding an output density function together with the value $\gamma > 0$ such that the above inequality is satisfied with equality on all $x \in X$ in the support of the capacity-achieving measure.

*Case $\eta > 2$:* We note that (8) has a constant part and a part that depends on $x$. It can be inspected that for large values of $x$ the term $\gamma \|x\|_\eta^\eta$ is a dominant term. Hence, for large values of $x$ to be in the support of $P_o$, the integral term must have a growth rate of $\|x\|_\eta^\eta$; otherwise, the support is bounded. As a result, it suffices to study the asymptotic behavior (tail) of the density function $f_{P_o, Q_{\bar{H}}}(y) = \frac{d(P_o W_{Q_{\bar{H}}})}{d\mu_Y}$. We have

$$f_{P_o, Q_{\bar{H}}}(y) = \frac{1}{\pi^m \sigma^{2m}} \int e^{-\frac{\|y - \bar{H}x\|_2^2}{\sigma^2}} dP_o$$

$$\geq \frac{1}{\pi^m \sigma^{2m}} \int e^{-\frac{\|y - \bar{H}x\|_2^2}{\sigma^2}} e^{-\frac{\|y + \bar{H}x\|_2^2}{\sigma^2}} dP_o$$

$$= e^{-\frac{2\|y\|_2^2}{\sigma^2}} \frac{1}{\pi^m \sigma^{2m}} \int e^{-\frac{2\|\bar{H}x\|_2^2}{\sigma^2}} dP_o.$$

This means that $-\log_2 f_{P_o, Q_{\bar{H}}}(y) = O(\|y\|_2^2)$. As a result, it can be verified that for $\eta > 2$, there exists no choice for the input measure so that the integral part of (8) catch up with the growth rate of $\|x\|_\eta^\eta$ for large values of $x$. Thus, the support of the capacity-achieving input measure is bounded for $\eta > 2$. One can verify that the hypothesis of Proposition 4.3 of Part I holds for Gaussian channels. More specifically, the function $\rho(z)$ is analytic on $Z$ except possibly at $z = 0$. As a result, we deduce that $\mathcal{S}_x(P_o)$ can not have any interior point.

*Case $\eta \leq 2$:* Note that every $y \in Y$ can be uniquely decomposed as $y = y_{\bar{H}} + y_{\bar{H}^\perp}$ where $y_{\bar{H}}$ is in the column space of $\bar{H}$ and $y_{\bar{H}^\perp}$ is orthogonal to the column space of $\bar{H}$. One can observe that the contribution of $y_{\bar{H}^\perp}$ is on the constant terms of (8) and it does not affect our analysis on the terms that depend on $x$. As a result, for every $y \in Y$ in the column space of $\bar{H}$, we have

$$f_{P_o, Q_{\bar{H}}}(y) = \frac{1}{\pi^m \sigma^{2m}} \int e^{-\frac{\|y - \bar{H}x\|_2^2}{\sigma^2}} dP_o$$

$$\geq \sup_r \frac{1}{\pi^m \sigma^{2m}} \int_{\|y - \bar{H}x\|_2 \leq r\sigma} e^{-\frac{\|y - \bar{H}x\|_2^2}{\sigma^2}} dP_o$$

$$\geq \frac{1}{\pi^m \sigma^{2m}} \sup_r e^{-r^2} P_o(\|y - \bar{H}x\|_2 \leq r\sigma)$$

Let $k(y) \triangleq \sup_r e^{-r^2} P_o(\|y - \bar{H}x\|_2 \leq r\sigma)$. Then, for every $y$ in the column space of $\bar{H}$, we deduce that

$$-\log_2 f_{P_o, Q_{\bar{H}}}(y) = O(\min(-\log_2 k(y), \|y\|_2^2)).$$

Two scenarios can be considered. Either the support is bounded or it is not. For the latter case to be true, we need to have the integral term of growth rate of $\|x\|_\eta^\eta$. Hence, it is necessary to have $k(y) = O(e^{-\alpha \|y\|_\eta^\eta})$ for sufficiently large $y$ in the column space of $\bar{H}$ where $\alpha$ is a positive real number. We remark that this necessary condition remains valid for the other case also. □

Proposition 3.3 provide us intuition about the support and the possible behavior of capacity achieving measures subject to different choices for $\eta$. However, it is still not possible to obtain closed form expressions for the the capacity-achieving measure for the general choice of $g(\cdot) = \|\cdot\|_\eta^\eta$ ($\eta \neq 2$). However, for the case of $\eta = 2$, it is possible to characterize the capacity achieving measure as shown in [8]. Using the observation that among distributions with the same covariance matrix, the Gaussian distribution achieves the largest relative entropy [23], Telatar [8] proves that the capacity-achieving input measure must have a Gaussian density function. Here, we use Kuhn-Tucker conditions, i.e., Theorem 4.3 of Part I, and provide a different approach toward characterizing the capacity-achieving input measure.

**Theorem 3.1.** *Let $g(\cdot) = \|\cdot\|_2^2$. The probability measure in $\mathscr{P}_{g,\Gamma}(X)$ which achieves the capacity of the channel $W_G(\bar{H})$ is absolutely continuous with respect to the Lebsegue measure with a unique (a.e.) Gaussian density function of zero mean and covariance matrix to*

$$S_x = \left[\frac{1}{\mu}I - \sigma^2(\bar{H}'\bar{H})^{-1}\right]^+,$$

*and $\mu$ is selected such that $\operatorname{tr}(S_x) = \Gamma$. Moreover, the capacity of this channel is*

$$C = \log_2 \det(I + \frac{1}{\sigma^2}\bar{H}S_x\bar{H}').$$

*Proof.* Suppose $P_o W_{Q_v}$ is the optimal capacity-achieving output measure. Applying Kuhn-Tucker condition, Theorem 4.3 of Part I, to our problem, we need to find a positive value $\gamma > 0$ such that

$$D(W_{Q_{\bar{H}}}(\cdot|x)\|P_o W_{Q_{\bar{H}}}) - \gamma\|x\|_2^2 \leq C - \gamma\Gamma,$$

for every $x \in X$. Using some straightforward mathematical manipulation, this results to

$$-m\log_2(\pi\sigma^2 e) - \int \log_2 \frac{d(P_o W_{Q_{\bar{H}}})}{d\mu_Y} \frac{1}{(\pi\sigma^2)^m} e^{-\frac{\|y-\bar{H}x\|_2^2}{\sigma^2}} d\mu_Y - \gamma\|x\|_2^2 \leq C - \gamma\Gamma. \qquad (9)$$

Now, the problem is to find an output density function together with the value $\gamma > 0$ such that the above inequality is satisfied with equality on all $x \in X$ in the support of the capacity-achieving measure.

Suppose $x \in X$ be in the support of the optimizing input measure. Then, we need to have the integration in the above inequality result into a quadratic form. To obtain a quadratic form out of integral, an straightforward option is to assume $\log_2 \frac{d(P_o W_{Q_v})}{d\mu_Y} = a - \|by\|_2^2$, where $a$ is a constant value and $b$ is an $m \times m$ matrix. As a result, we will obtain

$$-m\log_2(\pi\sigma^2 e) - a + \sigma^2 \operatorname{tr}(b'b) + \operatorname{tr}(x'\bar{H}'b'b\bar{H}x) - \gamma\|x\|_2^2 - C + \gamma\Gamma = 0$$

But to solve such equation, we can separate the constant part and the variable part to obtain

$$\begin{cases} x'(\bar{H}'b'b\bar{H} - \gamma I)x = 0 \\ -m\log_2(\pi\sigma^2 e) - a + \sigma^2 \operatorname{tr}(b'b) - C + \gamma\Gamma = 0 \end{cases} \qquad (10)$$

To satisfy the first equation of (10), we need to consider two cases. For any $x$ in the support of $P$, it suffices to have $x$ as an eigenvector of $\bar{H}'b'b\bar{H}$ with its corresponding eigenvalue equal to $\gamma$. This implies that we can take $b$ such that $\bar{H}'b'b\bar{H}$ represents a projection matrix such that its column space denotes the set of all $x$ in the support of $P_o$. For any $x$ not in the support, it suffices to have $x'(\bar{H}'b'b\bar{H} - \gamma I)x \leq 0$. Let $t$ denote the dimension of the space that is spanned by the vectors in support. Let $M_t$ denote a diagonal $m \times m$ matrix with the first $t$ elements equal to 1, and the rest of them 0. Thus, it suffices to take $b$ such that $b'b = \gamma(\bar{H}\bar{H}')^\dagger M_t + \beta(I - M_t)$, where $(\cdot)^\dagger$ denotes the pseudo-inverse operator and $\beta > 0$ is selected later to satisfy the above requirements. Note that $t$ is smaller or equal to the rank of $H$, since if some $x$ is in the null space of $\bar{H}$, it can not be in the support. Thus, $b'b$ is a full-rank positive definite matrix. This implies that the density function of $P_o W_{Q_v}$ with respect to the Lebesgue measure is Gaussian with zero mean and covariance $\frac{\log_2 e}{2}\left(\gamma^{-1}\bar{H}\bar{H}'M_t + \beta^{-1}(I - M_t)\right)$. But, to have such Gaussian density at the output, it suffices to have a Gaussian input with zero mean and covariance $S_x$ such that

$$\log_2 e \left(\gamma^{-1}\bar{H}\bar{H}'M_t + \beta^{-1}(I - M_t)\right) = \bar{H}S_x\bar{H}' + \sigma^2 I.$$

Thus, we need to pick $S_x$, $\gamma$, $\beta$, and $t$ to satisfy the above requirement. By Kuhn-Tucker condition we need to have $\gamma(\int \|x\|_2^2 dP_o - \Gamma) = 0$. Since, $\gamma$ can not be zero, we need to find $\gamma$ in the late equation with the constraint $\mathrm{tr}\,(S_x) = \Gamma$. As the result, it suffices to take $\beta = \frac{\log_2 e}{\sigma^2}$, and to take

$$S = \left[\frac{\log_2 e}{\gamma}I - \sigma^2(\bar{H}'\bar{H})^\dagger\right]^+ M_t,$$

where $[\cdot]^+$ ceils negative eigenvalues to 0. It just remains to select $t$, $\gamma$ such that $\mathrm{tr}\,(S) = \Gamma$. We emphasize that $t \leq \mathrm{rank}(\bar{H})$. Thus, to find the solution we first pick $t = \mathrm{rank}(\bar{H})$ and search for the value of $\gamma$. If we find the solution, we stop. Otherwise we reduce $t$ by 1 and repeat the procedure till we find the solution.

Picking $\gamma$ and $S$ as explained would result to obtain

$$a = -m\log_2(\pi) - \log\det\left(\frac{\log_2 e}{\gamma}\bar{H}\bar{H}'M_t + \sigma^2(I - M_t)\right).$$

Furthermore, we would have $\sigma^2 \mathrm{tr}\,(b'b) = m\log_2 e - \gamma\Gamma$. Thus, substituting and satisfying the equality in the second equation of (10) would result to

$$C = \log_2 \det(I + \frac{1}{\sigma^2}\bar{H}S\bar{H}'),$$

where $S$ is chosen as explained. For the sake of convenience in presentation, let $\mu = \frac{\gamma}{\log_2 e}$,

then we can simplify determination of $S$ to

$$S = \left[\frac{1}{\mu}I - \sigma^2(\bar{H}'\bar{H})^{-1}\right]^+,$$

where $\mu$ is selected such that $\text{tr}(S) = \Gamma$.

The uniqueness property follows by Lemma 3.2 and the fact that that there exists no other choice for $P_o$ to yield to the same Gaussian density function on the output. $\square$

## 4  Full Channel State Information at the Receiver

In mobile communications, sometimes it is possible to obtain an estimation of the channel realization at the receiver. This is specifically true in block-fading channels with a large coherence time, i.e., where there is sufficient delay between changes in channel realizations. In these systems, the transmitter assigns a portion of each block for training, where it sends some known signals to the receiver, so that the receiver obtains an estimate of the current channel realization. This process, called as *channel estimation*, allows the receiver to obtain some information about the channel state from none to (asymptotically) full channel state information (CSI).

In this section, we assume that full CSI is available at the receiver. Considering the general linear statistical model (1), this means that the channel realization, $H_k$, changes through the time but it is perfectly known at the receiver. We assume that the channel is Rayleigh or Rician faded [24], [5]. This means that the channel realization changes in accordance of a probability measure with a Gaussian density function with respect to the Lebesgue measure. If the density function is centralized (zero mean), the fading is Rayleigh; otherwise, it is Rician. In either case, the channel statistical law is fully characterized by the line-of-sight (mean) and scattering component (covariance matrix) of the channel realization [24], [5].

Let $\text{vec}(\cdot)$ denotes the *vector operator* that concatenates the columns of an $m \times n$ matrix, respectively, into an $mn \times 1$ vector. Thus, for every channel realization $H$, we denote its vector form as $\text{vec}\,H$, which is a multivariate random vector in $\mathbb{C}^{mn}$ that is characterized by the mean value vec $\bar{H}$ and the spatial covariance matrix

$$\Sigma = \text{cov}(\text{vec } H, \text{vec } H).$$

To emphasize that the channel state information is fully known, we use a subscript "$F$". As a result, we denote the channel as $W_F(\bar{H}, \Sigma)$. Since the channel realization is known at the receiver, there exists no advantage in taking $L > 1$ in the capacity analysis. Hence, likewise the previous section, we assume $L = 1$.

Since the channel realization is fully known at the receiver, we assume that $V = \mathbb{C}^{m \times n}$ is the space of state information and the conditional probability measure $Q_v$ on $\boldsymbol{H}$ is a point mass measure at $H = v$. Since the channel is Rayleigh or Rician fading, the probability measure on $V$, i.e., $R$, is absolutely continuous with respect to the Lebesgue measure with the Gaussian density function

$$\frac{dR}{d\mu_V} = \frac{1}{\pi^{mn} \det \Sigma} e^{-\text{vec}'(v-\bar{H})\Sigma^{-1}\text{vec}(v-\bar{H})}. \tag{11}$$

By definition of the auxiliary measure $T$ (3), one can inspect that $W_{Q_v}(\cdot|x) \ll T$ for every $v$ and $x$, where its density function is obtained from (4) as

$$f_{T,Q_v}(y|x) = \int f_T(y|x, H) dQ_v = \beta e^{\frac{\alpha^2 \|y\|_2^p - \|y-vx\|_2^2}{\sigma^2}}. \tag{12}$$

For every $P \in \mathscr{P}_{g,\Gamma}(X)$, it can be verified that $PW_{Q_v} \ll T$ with the density function

$$f_{T,P,Q_v}(y) = \beta \int e^{\frac{\alpha^2 \|y\|_2^p - \|y-vx\|_2^2}{\sigma^2}} dP. \tag{13}$$

As a result, the mutual information of this channel can be expressed as

$$I(P, W_F(\bar{H}, \Sigma)) = \iiint f_{T,Q_v}(y|x) \log_2 f_{T,Q_v}(y|x) dT dP dR$$
$$- \iint f_{T,P,Q_v}(y) \log_2 f_{T,P,Q_v}(y) dT dR. \tag{14}$$

## 4.1 Properties of mutual information

In this subsection, we address properties such as strict concavity and continuity of the mutual information, which are used in capacity analysis of fading channels with full CSI at the receiver.

In Proposition 3.3 of Part I, we have addressed the strict concavity of the mutual information. Since this property is essential to addressing the uniqueness of the capacity-achieving probability measure, we address a simpler condition to verify the strict concavity of the mutual information for fading channels with full CSI at the receiver. Using the following observation, we address this issue.

**Observation 4.1.** *For every $P \in \mathscr{P}_{g,\Gamma}(X)$, $f_{T,P,Q_v}(y)$ is continuous and nonzero over $Y \times V$.*

*Proof.* By (12), the continuity and positiveness of $f_{T,Q_v}(y|x)$ are obvious. The positiveness of $f_{T,Q_v}(y|x)$ implies the positiveness of $f_{T,P,Q_v}(y)$. To prove the continuity, suppose that $(y_n, v_n) \to (y, v)$ in the Euclidean norm. Then,

$$\lim_n f_{T,P,Q_{v_n}}(y_n) = \beta \lim_n \int e^{\frac{\alpha^2 \|y_n\|_2^P - \|y_n - v_n x\|_2^2}{\sigma^2}} dP$$

$$= \beta \lim_n e^{\frac{\alpha^2 \|y_n\|_2^P}{\sigma^2}} \lim_n \int e^{-\frac{\|y_n - v_n x\|_2^2}{\sigma^2}} dP$$

$$\text{By DCT} = \beta e^{\frac{\alpha^2 \|y\|_2^P}{\sigma^2}} \int e^{-\frac{\|y - \bar{H}x\|_2^2}{\sigma^2}} dP$$

$$= f_{T,P,Q_v}(y).$$

This proves the continuity of $f_{T,P,Q_v}(y)$ over $Y \times V$. □

We define that two probability measures $P_1, P_2 \in \mathscr{P}_{g,\Gamma}(X)$ are *equivalent over* $W_F(\bar{H}, \Sigma)$, if they induce the same conditional output probability measure (conditional on $v$). Equivalently, two input measures are equivalent over $W_F(\bar{H}, \Sigma)$ if $f_{T,P_1,Q_v}(y) = f_{T,P_2,Q_v}(y)$ for all $(y, v) \in Y \times V$ such that $v$ is in the support of $R$. Note that for full-rank $\Sigma$, all $v \in V$ are in the support of $R$.

**Proposition 4.1 (Strict concavity).** *The mutual information of channel $W_F(\bar{H}, \Sigma)$ is strictly concave with respect to the convex combination of two input measures $P_1$ and $P_2$, unless they are equivalent on $W_F(\bar{H}, \Sigma)$.*

*Proof.* By Observation 4.1, if there exists $(y, v) \in Y \times V$ such that $f_{T,P_1,Q_v}(y) \neq f_{T,P_2,Q_v}(y) \neq 0$ then there exists a neighborhood $U_{(y,v)} \subset Y \times V$ of $(y, v)$ with that property. By definition of $T$ (3) and $R$ (11), $(T \times R)(U_{(y,v)}) > 0$. This implies that the set $U_{(y,v)}$ complies the requirements of Proposition 3.3 of Part I. Thus, the mutual information function of an FCSI channel is strictly concave with respect to the convex combination of $P_1$ and $P_2$ if and only if there exists $(y, v) \in Y \times V$ such that $f_{T,P_1,Q_v}(y) \neq f_{T,P_2,Q_v}(y)$. □

Another important property of the mutual information in capacity analysis is its weak* continuity, which is stated as follows.

**Proposition 4.2 (Continuity).** *The mutual information of any fading channel $W_F(\bar{H}, \Sigma)$ is weak\* continuous over $\mathscr{P}_{g,\Gamma}(X)$.*

*Proof.* It suffices to check if the hypothesis of Theorem 3.3 of Part I is satisfied. We prove the theorem for $g(x) = \|x\|_2^2$, and the proof for other choices of $g$ is also similar. To prove hypothesis (a) of Theorem 3.3 of Part I, we proceed as the proof of Proposition 3.2.

Let assume that $\alpha < 1$ and $p < 2$ are fixed. Let define

$$A_c = \{(x, y, v) \in X \times Y \times V | f_{T,Q_v}(y|x) |\log_2 f_{T,Q_v}(y|x)| > c^2\}.$$

Let also define

$$B_c = \{(x, y, v) \in X \times Y \times V | f_{T,Q_v}(y|x) > c\}.$$

For sufficiently large $c > 0$, e.g. $c > 10$, it can be observed that $A_c \subset B_c$. But $f_{T,Q_v}(y|x) > c$ is equivalent to $\alpha^2 \|y\|_2^p - \|y - vx\|_2^2 > \sigma^2 \ln \frac{c}{\beta}$. Hence, if we define

$$E_c = \{(x, y, v) \in X \times Y \times V | \alpha^2 \|y\|_2^p > \sigma^2 \ln \frac{c}{\beta}\},$$

we can deduce that for sufficiently large $c > 0$, we would have $A_c \subseteq B_c \subseteq E_c$. Let $\omega(c) = \left(\frac{\sigma^2}{\alpha^2} \ln \frac{c}{\beta}\right)^{1/p}$. We have,

$$\iiint_{A_c} f_{T,Q_v}(y|x) |\log_2 f_{T,Q_v}(y|x)| dT dP dR \leq \iiint_{E_c} f_{T,Q_v}(y|x) |\log_2 f_{T,Q_v}(y|x)| dT dP dR$$

$$\leq |\log_2 \beta| \iiint_{E_c} \frac{1}{\pi^m \sigma^{2m}} e^{-\frac{\|y-vx\|_2^2}{\sigma^2}} d\mu_Y dP dR$$

$$+ \frac{\log_2 e}{\sigma^2} \iiint_{E_c} \frac{\alpha^2 \|y\|_2^p}{\pi^m \sigma^{2m}} e^{-\frac{\|y-vx\|_2^2}{\sigma^2}} d\mu_Y dP dR$$

$$\leq \left(\frac{|\log_2 \beta|}{\omega^2(c)} + \frac{\log_2 e}{\sigma^2 \omega^{2-p}(c)}\right)$$

$$\iiint \frac{\|y\|_2^2}{\pi^m \sigma^{2m}} e^{-\frac{\|y-vx\|_2^2}{\sigma^2}} d\mu_Y dP dR$$

$$= \left(\frac{|\log_2 \beta|}{\omega^2(c)} + \frac{\log_2 e}{\sigma^2 \omega^{2-p}(c)}\right) \left(m\sigma^2 + \iint \|vx\|_2^2 dP dR\right)$$

$$\leq \left(\frac{|\log_2 \beta|}{\omega^2(c)} + \frac{\log_2 e}{\sigma^2 \omega^{2-p}(c)}\right) \left(m\sigma^2 + \Gamma \int \|v\|^2 dR\right).$$

Since $(m\sigma^2 + \Gamma \int \|v\|^2 dR) < \infty$ and $\omega(c) \to \infty$ as $c \to \infty$, we can deduce that

$$\lim_{c \to \infty} \sup_{P \in \mathscr{P}_{g,\Gamma}(X)} \iiint_{A_c} f_{T,Q_v}(y|x) |\log_2 f_{T,Q_v}(y|x)| \, dT dP dR = 0.$$

This implies that hypothesis (a) of Theorem 3.3 of Part I holds.

To verify if hypothesis (b) holds, for fixed $y$ and $v$, let $B_c = \{x \in X | f_{T,Q_v}(y|x) > c\}$. Then,

$$\sup_{P \in \mathscr{P}_{g,\Gamma}(X)} \int_{B_c} f_{T,Q_v}(y|x) dP \leq \frac{1}{c} \sup_{P \in \mathscr{P}_{g,\Gamma}(X)} \int (f_{T,Q_v}(y|x))^2 dP$$

$$= \frac{\beta^2}{c} \sup_{P \in \mathscr{P}_{g,\Gamma}(X)} \int e^{2\frac{\alpha^2 \|y\|_2^p - \|y - vx\|_2^2}{\sigma^2}} dP$$

$$\leq \frac{\beta^2}{c} e^{2\frac{\alpha^2 \|y\|_2^p}{\sigma^2}} \to 0, \text{ as } c \to \infty.$$

Thus, both hypotheses of Theorem 3.3 of Part I hold. Hence, the mutual information is weak* continuous. □

## 4.2 Capacity analysis

In this subsection, we address issues on the existence, the uniqueness, and the characterization of the capacity-achieving measure for fading channels with full CSI at the receiver.

**Lemma 4.1 (Existence).** *For any channel $W_F(\bar{H}, \Sigma)$ and the set of input measures $\mathscr{P}_{g,\Gamma}(X)$, there exists a measure $P_o \in \mathscr{P}_{g,\Gamma}(X)$ that achieves the capacity of $W_F(\bar{H}, \Sigma)$.*

*Proof.* By Proposition 4.2, the mutual information is continuous over $\mathscr{P}_{g,\Gamma}(X)$. Since $\mathscr{P}_{g,\Gamma}(X)$ is weak* compact, the existence is guaranteed by Proposition 4.1 of Part I. □

One immediate result of our arguments on strict concavity of mutual information, Proposition 4.1, is on the uniqueness of the capacity-achieving measure, as shown below.

**Lemma 4.2 (Uniqueness).** *For any fading channel $W_F(\bar{H}, \Sigma)$ and the set of input measures $\mathscr{P}_{g,\Gamma}(X)$, the capacity-achieving measure is unique up to the equivalency of input measures.*

*Proof.* The proof is similar to the proof of Lemma 3.2. □

We remark that using Lemma 4.2 with some simple intuitive arguments, one can justify that the capacity-achieving input measure is symmetric.

So far, in this section, we have addressed issues on the existence and the uniqueness of the capacity-achieving measure over fading channels with full CSI at the receiver. It remains to provide some insight to the characterization of the capacity-achieving measure.

**Proposition 4.3.** *Let $g(\cdot) = \|\cdot\|_\eta^\eta$ for $1 \leq \eta$. If $\eta > 2$, the capacity-achieving measure has a bounded support with no interior point. In contrast, if $1 \leq \eta \leq 2$, then a necessary condition for $P_o$ is that for almost every $v \in V$, $\sup_{r \geq 0} e^{-r^2} P_o(\|y - vx\|_2 \leq r\sigma) = O(e^{-\alpha(v)\|y\|_\eta^\eta})$ for sufficiently large $y$ in the column space of $v$ and for some positive function $\alpha : V \to \mathbb{R}^+$.*

*Proof.* Let $P_o W_{Q_v}$ denote the optimal conditional capacity-achieving output measure. Applying Kuhn-Tucker condition, Theorem 4.3 of Part I, to our problem, we need to find a positive value $\gamma > 0$ such that
$$\int D(W_{Q_v}(\cdot|x) \| P_o W_{Q_v}) dR - \gamma \|x\|_\eta^\eta \leq C - \gamma \Gamma.$$
Using some straightforward mathematical manipulation, this results to

$$-m \log_2(\pi \sigma^2 e) - \iint \frac{1}{(\pi\sigma^2)^m} e^{-\frac{\|y-vx\|_2^2}{\sigma^2}} \log_2 f_{P_o, Q_v}(y) d\mu_Y dR - \gamma \|x\|_\eta^\eta \leq C - \gamma \Gamma. \quad (15)$$

The problem is now finding an output density function together with the value $\gamma > 0$ such that the above inequality is satisfied with equality on all $x \in X$ in the support of the capacity-achieving measure.

It can be inspected that for large values of $x$ the term $\gamma \|x\|_\eta^\eta$ is a dominant term. Hence, the integral term must have a growth rate equal or smaller than $\gamma \|x\|_\eta^\eta$. As a result, it suffices to study the asymptotic behavior (tail) of the density function $f_{P_o, Q_v}(y) = \frac{d(P_o W_{Q_v})}{d\mu_Y}$.

Suppose $y \in Y$ and $v \in V$ are fixed and $y$ is in the column space of $v$. Let $k_v(y) = \sup_r e^{-r^2} P_o(\|y - vx\|_2 \leq r\sigma)$. Similar to the proof of Proposition (3.3), we can deduce that for $y$ in the column space of $v$,
$$-\log_2 f_{P_o, Q_v}(y) = O(\min(-\log_2 k_v(y), \|y\|_2^2)).$$

Now, let consider this together with (15). It can be verified that for $\eta > 2$, there exists no choice for the input measure to catch up with the growth rate of $\|x\|_\eta^\eta$ for large values of $x$. This implies that the support of the input measure is bounded for $\eta > 2$. One can verify that the hypothesis of Proposition 4.3 of Part I holds for fading channels with full CSI. More specifically, the function $\rho(z)$ is analytic on $Z$ except possibly at $z = 0$. As a result, we deduce that $\mathcal{S}_x(P_o)$ can not have any interior point. On the other hand, for $\eta \leq 2$, a necessary condition for the input measure is $k(y) = O(e^{-\alpha(v)\|y\|_\eta^\eta})$ for sufficiently large $y$ in the column space of $v$ where $\alpha : V \to \mathbb{R}^+$. □

The capacity-achieving measures for the general choice of $g(\cdot) = \|\cdot\|_\eta^\eta$ are not known. However, for the case that $\eta = 2$, this problem was first addressed in [8] and solved for i.i.d Rayleigh distribution. Telatar [8] showed that the capacity-achieving input distribution is an isotropic Gaussian distribution. Foschini [7] has shown similar results also.

Here, we want to address the capacity of the channel in the presence of arbitrary correlation and line-of-sight components. For this purpose, we use Kuhn-Tucker condition, Theorem 4.3 of Part I.

**Theorem 4.1.** *Let $g(\cdot) = \|\cdot\|_2^2$. The probability measure in $\mathscr{P}_{g,\Gamma}(X)$ that achieves the capacity of the channel $W_F(\bar{H}, \Sigma)$ is absolutely continuous with respect to the Lebesgue measure with a unique (a.e.) Gaussian density function of zero mean and covariance matrix $S$ that satisfies*

$$\int x'v'(vSv' + \sigma^2 I_m)^{-1}vx\,dR \leq \mu\|x\|_2^2, \ \forall x \in X$$

*where the equality occurs if and only if $x$ is in the support of capacity-achieving measure and $\mu > 0$ is selected to satisfy $\operatorname{tr}(S) = \Gamma$. Moreover, the capacity of this channel is*

$$C = \int \log\det(I_m + 1/\sigma^2 vSv')\,dR.$$

*Proof.* Suppose $P_o W_{Q_v}$ is the optimal capacity-achieving output measure. Applying Kuhn-Tucker condition, Theorem 4.3 of Part I, to our problem, we need to find a positive value $\gamma > 0$ such that

$$\int D(W_{Q_v}(\cdot|x)\|P_o W_{Q_v})\,dR - \gamma\|x\|_2^2 \leq C - \gamma\Gamma.$$

Using some straightforward mathematical manipulation, this results to

$$-m\log_2(\pi\sigma^2 e) - \iint \log_2 \frac{d(P_o W_{Q_v})}{d\mu_Y} \frac{1}{(\pi\sigma^2)^m} e^{-\frac{\|y-vx\|_2^2}{\sigma^2}} d\mu_Y dR - \gamma\|x\|_2^2 \leq C - \gamma\Gamma.$$

The problem is now finding an output density function together with the value $\gamma > 0$ such that the above inequality is satisfied with equality on all $x \in X$ in the support of the capacity-achieving measure.

Suppose $x \in X$ be in the support of the optimizing input measure. Then, we need to have the integration in the above inequality result into a quadratic form. To obtain a quadratic form out of integral, an straightforward option is to assume $\log_2 \frac{d(P_o W_{Q_v})}{d\mu_Y} = a(v) - \|b(v)y\|_2^2$, where $a(v)$ is a function and $b(v)$ is a mapping from $V$ to the $M_m(\mathbb{R})$. As the result, we will obtain

$$-m\log_2(\pi\sigma^2 e) + \int [-a(v) + \sigma^2 \mathrm{tr}\,(b(v)'b(v)) + x'v'b(v)'b(v)vx] dR - \gamma\|x\|_2^2 - C + \gamma\Gamma = 0$$

But to solve such equation, we can separate the constant part and the variable part to obtain

$$\begin{cases} \int x'v'b(v)'b(v)vx dR - \gamma\|x\|_2^2 = 0 \\ -m\log_2(\pi\sigma^2 e) + \int [-a(v) + \sigma^2 \mathrm{tr}\,(b(v)'b(v))] dR - C + \gamma\Gamma = 0 \end{cases} \quad (16)$$

To satisfy the first equation of (16), we proceed as follows. For any $x$ in the support of $P_o$, it suffices to have $x$ as an eigenvector of $\mathrm{E}\,(v'b(v)'b(v)v)$ with its corresponding eigenvalue equal to $\gamma$. For any $x$ not in the support of $P_o$, we must have $x'\mathrm{E}\,(v'b(v)'b(v)v)\,x - \gamma\|x\|_2^2 < 0$. This implies that we should select $b(v)$ such that the maximal eigenvalues of $\mathrm{E}\,(v'b(v)'b(v)v)$ (that correspond to the support of $P_o$ be $\gamma$) and the rest of its eigenvalues be less than $\gamma$. One immediate approach to select $b(v)$ is to assume that the input measure is a centralized multivariate normal with covariance $S$. As a result, this implies that $b(v)'b(v) = \log_2 e\,(vSv' + \sigma^2 I_m)^{-1}$. Therefore, it remains just to pick a semi-positive definite matrix $S$ such that $\log_2 e\,\mathrm{E}\,(v'(vSv' + \sigma^2 I_m)^{-1}v)$ have our desired structure. That is we need to find $S$ such that

$$\log_2 e\,\mathrm{E}\,(v'(vSv' + \sigma^2 I_m)^{-1}v) - \gamma I \leq 0$$

in consideration of other constraints raising from (16). It can be inspected that such choice for $S$ exists and depends on the value of $\gamma$. By Kuhn-Tucker condition we need to have $\gamma(\int \|x\|_2^2 dP_o - \Gamma) = 0$. Since $\gamma$ can not be zero, we need to find $\gamma$ such that $\mathrm{tr}\,(S) = \Gamma$. For convenience in presentation, let define $\mu = \frac{g}{\log_2 e}$. Now, multiplying the above equation from left by $x'$ and right by $x$, and taking the expectation with respect to $x$, one can verify that we obtain the equality

$$\int \sigma^2 \text{tr}\left((vSv' + \sigma^2 I_m)^{-1}\right) dR - m\log_2 e + \mu\Gamma = 0 \qquad (17)$$

Since we have

$$a(v) = -m\log_2 \pi - \frac{1}{2}\log\det(vSv' + \sigma^2 I_m).$$

Substituting in (16), we obtain

$$\int \log_2 \det(I_m + 1/\sigma^2 vSv')dR - C + \int \sigma^2 \text{tr}(b(v)'b(v)) dR - m\log_2 e + \mu\Gamma = 0$$

In consideration of (17), we would obtain

$$C = \int \log_2 \det(I_m + 1/\sigma^2 vSv')dR.$$

The uniqueness property follows by from Lemma 4.2 and the fact that there exists no other input measure than can induce the same conditional output measure with Gaussian density function. □

Note that Theorem 4.1 characterizes the capacity-achieving measure for any Rician MIMO channel with full CSI at the receiver. In general, a closed form solution for $S$ is not obtainable from Theorem 4.1, and $S$ should be found through exhaustive computer search. However, for special cases of channels, one may use some novel approaches to solve the conditions in Theorem 4.1. As an example, consider the following corollary.

**Corollary 4.1.** *If $W_F(\bar{H}, \Sigma)$ is an i.i.d. Rayleigh channel, i.e., $\bar{H} = 0$ and $\Sigma = I$, then the capacity-achieving measure has an isotropic Gaussian density function with zero mean and covariance matrix $S = \frac{\Gamma}{n}I$.*

*Proof.* Suppose $S$ is the covariance matrix that satisfies the condition in Theorem 4.1. Let $U$ be an $n \times n$ unitary matrix. Since $R$ is invariant under the operation of $U$, it can be inspected that $USU'$ is also a right candidate. By uniqueness of the capacity-achieving measure, we deduce that $S = USU'$ for every unitary matrix $U$. This implies that $S = \lambda I$ for some $\lambda > 0$. Since $\text{tr}(S) = \Gamma$, we deduce that $S = \frac{\Gamma}{n}I$. □

Recall that this is the same result as given in [8] for Rayleigh channels. One can verify that if $\Sigma$ is not singular, then for large signal-to-noise ratio (SNR), i.e. $\frac{\Gamma}{\sigma^2}$, the covariance matrix $S$ would be very close to $\frac{\Gamma}{n}I$. Hence, for large SNR, an input measure with isotropic Gaussian distribution is near optimal. Relevant work can be found in [9], [25], and [26].

## 5 No Channel State Information at the Receiver

In some applications, the system setup does not allow any estimation of the channel realization at the receiver. This is specifically true in fast fading channels, where there is not sufficient delay available between changes in the channel realization.

In this section, we consider Rayleigh or Rician fading channels where no CSI is available at the receiver. Regarding the general linear statistical model as (1), this means that the channel realization, $H_k$, changes through time in accordance of a Gaussian probability density function, but the realization is not known at the receiver. As in the previous section, we characterize these channels with the mean value $\bar{H}$ and the covariance matrix $\Sigma$ of the fading. To emphasize that CSI is not known, we use a subscript "$N$", and denote the channel as $W_N(\bar{H}, \Sigma)$.

Since we assumed that no CSI is available at the receiver, the space of side information, $V$, is statistically independent from $\boldsymbol{H}$. In the framework of the generic system model (Section 3), this is equivalent to assume that an arbitrary Borel measurable space $V$ with an arbitrary measure $R$ on it such for every given $v$, the conditional probability measure on $\boldsymbol{H}$, $Q_v$, has a Gaussian density function characterized by $\bar{H}$ and $\Sigma$. Since the measure $Q_v$ is not dependent on $v$, we drop indexing by $v$ and simply use $Q$, instead.

Since the channel realization is not known at the receiver, the size of block length $L$ is important in capacity analysis of these channels. Hence, we assume $X = \mathbb{C}^{n \times L}$ for a general $L \geq 1$. By definition of the auxiliary measure $T$ (3), one can inspect that $W_Q(\cdot|x) \ll T$ for every $x$ with the density function (obtained from (4))

$$f_{T,Q}(y|x) = \frac{\beta}{\det \Phi_x} \exp\left(\frac{\alpha^2 \|y\|_2^p - \text{vec}\,(y - \bar{H}x)' \Phi_x^{-1} \text{vec}\,(y - \bar{H}x)}{\sigma^2}\right) \tag{18}$$

where $\Phi_x \triangleq \frac{1}{\sigma^2}(x' \otimes I_m)\Sigma(x \otimes I_m) + I_{mL}$. For every $P \in \mathscr{P}_{g,\Gamma}(X)$, let

$$f_{T,P,Q}(y) = \int f_{T,Q}(y|x) dP \tag{19}$$

denote the output density function. The mutual information of this channel is expressed as

$$I(P, W_N(\bar{H}, \Sigma)) = \iint f_{T,Q}(y|x) \log_2 f_{T,Q}(y|x) dT dP - \int f_{T,P,Q}(y) \log_2 f_{T,P,Q}(y) dT. \quad (20)$$

One should note that the maximum of (20) should be divided by $L$ to determine the capacity per channel use.

## 5.1 Properties of mutual information

In this subsection, we address properties such as strict concavity and continuity of the mutual information for fading channels with no CSI at the receiver.

Propositions 3.3 of Part I addressed the strict concavity of the mutual information in general. Since the strictness property is essential to addressing the uniqueness of the capacity-achieving probability measure, we address a simpler condition for the case of fading channels with no CSI at the receiver.

**Observation 5.1.** *For every $P \in \mathscr{P}_{g,\Gamma}(X)$, $f_{T,P,Q}(y)$ is continuous and nonzero over $Y$.*

*Proof.* By (18), the continuity and positiveness of $f_{T,Q}(y|x)$ is obvious. The positiveness of $f_{T,Q}(y|x)$ implies the positiveness of $f_{T,P,Q}(y)$. To prove the continuity, suppose that $y_n \to y$ in the Euclidean norm. Then,

$$\begin{aligned} \lim_n f_{T,P,Q}(y_n) &= \lim_n \int f_Q(y|x) dP \\ &= \lim_n \frac{\beta}{\det \Phi_x} e^{\frac{\alpha^2 \|y_n\|_2^p}{\sigma^2}} \lim_n \int \exp\left(\frac{-\text{vec}(y_n - \bar{H}x)' \Phi_x^{-1} \text{vec}(y_n - \bar{H}x)}{\sigma^2}\right) dP \\ \text{By DCT} &= \frac{\beta}{\det \Phi_x} e^{\frac{\alpha^2 \|y\|_2^p}{\sigma^2}} \int \exp\left(\frac{-\text{vec}(y - \bar{H}x)' \Phi_x^{-1} \text{vec}(y - \bar{H}x)}{\sigma^2}\right) dP \\ &= f_{T,P,Q}(y). \end{aligned}$$

This proves the continuity of $f_{T,P,Q}(y)$ over $Y$. □

We define that two probability measures $P_1, P_2 \in \mathscr{P}_{g,\Gamma}(X)$ are *equivalent over $W_N(\bar{H}, \Sigma)$* if they induce the same output probability measure. Equivalently, two input measures are equivalent over $W_N(\bar{H}, \Sigma)$ if $f_{T,P_1,Q}(y) = f_{T,P_2,Q}(y)$ for all $y \in Y$.

**Proposition 5.1 (Strict concavity).** *The mutual information of channel $W_N(\bar{H}, \Sigma)$ is strictly concave with respect to the convex combination of two input measures $P_1$ and $P_2$, unless they are equivalent.*

*Proof.* By Observation 5.1, if there exists $y \in Y$ such that $f_{T,P_1,Q}(y) \neq f_{T,P_2,Q}(y)$ then there exists a neighborhood $U_y \subset Y$ of $y$ with that property. Then by definition of $T$ (3), $T(U_y) > 0$. This implies that the set $U_y$ complies the requirements of Proposition 3.3 of Part I. Thus, the mutual information of an NCSI channel is strictly concave with respect to the convex combination of $P_1$ and $P_2$ if and only if there exists $y \in Y$ such that $f_{T,P_1,Q}(y) \neq f_{T,P_2,Q}(y)$. By definition of equivalency of measures over $W_N(\bar{H}, \Sigma)$, we deduce the assertion. $\square$

In the following, we state and prove another important property of mutual information, weak* continuity, which will be used later in the capacity analysis of fading channels with no CSI.

**Proposition 5.2 (Continuity).** *The mutual information of any fading channel $W_N(\bar{H}, \Sigma)$ is weak* continuous over $\mathscr{P}_{g,\Gamma}(X)$.*

*Proof.* It suffices to check if the hypothesis of Theorem 3.3 of Part I is satisfied. We prove the theorem for $g(x) = \|x\|_2^2$, and the proof for other choices of $g$ is also similar. To prove hypothesis (a) of Theorem 3.3 of Part I, we proceed as the proof of Proposition 3.2. For every $P \in \mathscr{P}_{g,\Gamma}(X)$, let

$$A_c = \{(x,y) \in X \times Y | f_{T,Q}(y|x) |\log_2 f_{T,Q}(y|x)| > c^2\}.$$

Let

$$B_c = \{(x,y) \in X \times Y | f_{T,Q}(y|x) > c\},$$

for sufficiently large $c > 0$, it can be observed that $A_c \subset B_c$. Let

$$D_c = \{(x,y) \in X \times Y | \alpha^2 \|y\|_2^p - \text{vec}\,(y - \bar{H}x)' \Phi_x^{-1} \text{vec}\,(y - \bar{H}x) > \sigma^2 \ln \frac{c \det \Phi_x}{\beta}\}.$$

It can be inspected that $B_c \subset D_c$. We also note that or sufficiently large $c > 0$, $\alpha^2 \|y\|_2^p \geq \text{vec}\,(y - \bar{H}x)' \Phi_x^{-1} \text{vec}\,(y - \bar{H}x)$. Let also define

$$E_c = \{(x,y) \in X \times Y | \|y\|_2^p > \frac{\sigma^2}{\alpha^2} \ln \frac{c \det \Phi_x}{\beta}\}.$$

Thus, for sufficiently large $c > 0$, $D_c \subseteq E_c$. Noting that $\det \Phi_x \geq 1$, let define $\omega(c) =$

$\left(\frac{\sigma^2}{\alpha^2} \ln \frac{c}{\beta}\right)^{1/p}$. In a similar discussion as in of Proposition 3.2, we can deduce that

$$\iint_{A_c} f_{T,Q}(y|x) |\log_2 f_{T,Q}(y|x)| \, dTdP$$

$$\leq \iint_{E_c} f_{T,Q}(y|x) |\log_2 f_{T,Q}(y|x)| \, dTdP$$

$$\leq |\log_2 \beta| \iint_{E_c} \frac{1}{\pi^{mL}\sigma^{2mL} \det \Phi_x} e^{-\frac{1}{\sigma^2}\text{vec}(y-\bar{H}x)'\Phi_x^{-1}\text{vec}(y-\bar{H}x)} d\mu_Y dP$$

$$+ \frac{\log_2 e}{\sigma^2} \iint_{E_c} \frac{2\alpha^2 \|y\|_2^p}{\pi^{mL}\sigma^{2mL} \det \Phi_x} e^{-\frac{1}{\sigma^2}\text{vec}(y-\bar{H}x)'\Phi_x^{-1}\text{vec}(y-\bar{H}x)} d\mu_Y dP$$

$$\leq \left[\frac{|\log_2 \beta|}{\omega^2(c)} + \frac{2\alpha^2 \log_2 e}{\sigma^2 \omega^{2-p}(c)}\right]$$

$$\iint \frac{\|y\|_2^2}{\pi^{mL}\sigma^{2mL} \det \Phi_x} e^{-\frac{1}{\sigma^2}\text{vec}(y-\bar{H}x)'\Phi_x^{-1}\text{vec}(y-\bar{H}x)} d\mu_Y dP$$

If we take the $\sup_{P \in \mathscr{P}_{g,\Gamma}(X)}$ of both sides, the second term of the RHS is a finite value. Now, applying $\lim_{c \to \infty}$ to both sides, we observe that $\omega(c) \to \infty$ as $c \to \infty$. Hence,

$$\lim_{c \to \infty} \sup_{P \in \mathscr{P}_{g,\Gamma}(X)} \iint_{A_c} f_{T,Q}(y|x) |\log_2 f_{T,Q}(y|x)| \, dTdP = 0.$$

This implies that the hypothesis (a) of Theorem 3.3 of Part I holds. To verify if the hypothesis (b) holds, the proof is by Chebychev's inequality which is essentially similar to the proof of Proposition 4.2. Thus, both hypotheses of Theorem 3.3 of Part I hold, so the mutual information is weak* continuous. □

## 5.2 Capacity analysis

In this subsection, we address issues on the existence, the uniqueness, and the characterization of the capacity-achieving measures for fading channels with no CSI at the receiver.

**Lemma 5.1 (Existence).** *For any channel $W_N(\bar{H}, \Sigma)$, there exists a measure $P_o \in \mathscr{P}_{g,\Gamma}(X)$ that achieves the capacity of $W_N(\bar{H}, \Sigma)$ over $\mathscr{P}_{g,\Gamma}(X)$.*

*Proof.* Proposition 5.2 states the weak* continuity of the mutual information over $\mathscr{P}_{g,\Gamma}(X)$. Since $\mathscr{P}_{g,\Gamma}(X)$ is weak* compact, by Proposition 4.1 of Part I, we conclude the assertion. □

Lemma 5.1 states the existence of the capacity-achieving measure over fading channels with no CSI. One immediate result of our arguments on strict concavity of mutual infor-mation, Proposition 5.1, is on the uniqueness of the capacity-achieving measure, as shown below.

**Lemma 5.2 (Uniqueness).** *For any channel $W_N(\bar{H}, \Sigma)$, the capacity-achieving measure over $\mathscr{P}_{g,\Gamma}(X)$ is unique up to the equivalency of input measures.*

*Proof.* The proof is essentially the same as the proof of Lemma 3.2. □

So far, in this section, we have shown the existence and uniqueness of the capacity-achieving measure over fading channels with no CSI at the receiver. It remains to provide some insight to the characterization of such input measure.

**Proposition 5.3.** *Suppose $g(\cdot) = \|\cdot\|_\eta^\eta$ for $1 \leq \eta$. If $\eta > 2$, the capacity-achieving measure has a bounded support with no interior point. In contrast, if $1 \leq \eta \leq 2$, a necessary condition for the capacity-achieving measure is*

$$P_o(\|y\|_2 \leq \|x\|_2, \mathcal{R}(y) \subseteq \mathcal{R}(x)) = O(e^{-\alpha\|y\|_\eta^\eta}),$$

*where $\mathcal{R}(\cdot)$ denote the row space of a matrix, and $\alpha > 0$ is a constant.*

*Proof.* Suppose $P_o W_Q$ is the optimal capacity-achieving output measure. Applying Kuhn-Tucker condition, Theorem 4.3 of Part I, to our problem, we need to find a positive value $\gamma > 0$ such that

$$D(W_Q(\cdot|x)\|P_o W_Q) - \gamma\|x\|_\eta^\eta \leq C - \gamma\Gamma.$$

We note that $W_Q(\cdot|x) \ll \mu_Y$ with density function

$$f_Q(y|x) = \frac{1}{\pi^{mL}\sigma^{2mL}\det\Phi_x} e^{-\frac{1}{\sigma^2}\text{vec}(y-\bar{H}x)'\Phi_x^{-1}\text{vec}(y-\bar{H}x)},$$

where $\Phi_x$ is defined as in (18). Using some straightforward mathematical manipulation, we obtain

$$-mL\log_2(\pi e\sigma^2) - \int \log_2 \frac{d(P_o W_Q)}{d\mu_Y} f_Q(y|x)d\mu_Y - \log_2(\det \Phi_x) - \gamma\|x\|_\eta^\eta \leq C - \gamma\Gamma. \quad (21)$$

Now, the problem is to find an output density function together with the value $\gamma > 0$ such that the above inequality is satisfied with equality on all $x \in X$ in the support of the capacity-achieving measure. Unfortunately, because of the inherent difficulties in this expression, it is not possible to find an analytic solution for the output density function. However, through some asymptotic analysis discussion we can obtain some intuition on characterization of the support of the capacity-achieving measure, as follows.

*Case $\eta > 2$:* It can be inspected that for large values of $x$, the term $\gamma \|x\|_\eta^\eta$ is a dominant term. Hence, the integral term must have a growth rate of $\|x\|_\eta^\eta$; otherwise, the support of the capacity-achieving measure is bounded. Thus, it suffices to study the asymptotic (tail) behavior of the density function $f_{P_o,Q}(y) = \frac{d(P_o W_Q)}{d\mu_Y}$. For every fixed $y \in Y$, we have

$$f_{P_o,Q}(y) = \int \frac{1}{\pi^{mL}\sigma^{2mL}\det \Phi_x} e^{-\frac{1}{\sigma^2}\text{vec}(y-\bar{H}x)'\Phi_x^{-1}\text{vec}(y-\bar{H}x)} dP_o$$

$$\geq \int \frac{1}{\pi^{mL}\sigma^{2mL}\det \Phi_x} e^{-\frac{2}{\sigma^2}\text{vec}\,y'\Phi_x^{-1}\text{vec}\,y} e^{-\frac{2}{\sigma^2}\text{vec}(\bar{H}x)'\Phi_x^{-1}\text{vec}(\bar{H}x)} dP_o$$

$$\geq e^{-\frac{2}{\sigma^2}\|y\|_2^2} \int \frac{1}{\pi^{mL}\sigma^{2mL}\det \Phi_x} e^{-\frac{2}{\sigma^2}\text{vec}(\bar{H}x)'\Phi_x^{-1}\text{vec}(\bar{H}x)} dP_o$$

This means that $-\log_2 f_{P_o,Q_{\bar{H}}}(y) = O(\|y\|_2^2)$. As a result, it can be verified that for $\eta > 2$, there exists no choice for the input measure so that the integral part of (21) catch up with the growth rate of $\|x\|_\eta^\eta$ for large values of $x$. Thus, the support of the capacity-achieving input measure is bounded for $\eta > 2$. Trying to use Proposition 4.3 of Part I, one can verify that $\rho(z)$ is analytic over $Z$ except the zero set of some polynomials, say $A \in Z$. Since the zero set of polynomials are closed sets including boundary points [27], the set $U = Z \backslash A$ is a connected set. Because of the existence of $\log_2(\det \Phi_x)$ in (21), we know that the zero set of $\rho(z)$ is not the solution of a polynomial. Thus, $\xi(\mathcal{S}_x(P_o)) \cap U$ is not empty. Now, applying Proposition 4.3 of Part I, we deduce that $\mathcal{S}_x(P_o)$ can not have any interior point.

*Case $\eta \leq 2$:* Recalling that $\Phi_x = \frac{1}{\sigma^2}(x' \otimes I_m)\Sigma(x \otimes I_m) + I_{mL}$, one can verify that for every $x \in X$,

$$(\frac{\lambda_{\min}}{\sigma^2} x'x + I_L) \otimes I_m \leq \Phi_x \leq (\frac{\lambda_{\max}}{\sigma^2}\|x\|_2^2 + 1)I_{mL},$$

where $\lambda_{\max}$ and $\lambda_{\min} > 0$ are the maximum and minimum eigenvalues of $\Sigma$. Let the operator $\mathcal{R}(\cdot)$ denote the row space of a matrix. Then,

$$f_{P_o,Q}(y) \geq \int \frac{1}{\pi^{mL}\sigma^{2mL}\det \Phi_x} e^{-\frac{2}{\sigma^2}\text{vec}\,y'\Phi_x^{-1}\text{vec}\,y} e^{-\frac{2}{\sigma^2}\text{vec}(\bar{H}x)'\Phi_x^{-1}\text{vec}(\bar{H}x)} dP_o$$

$$\geq \int \frac{1}{\pi^{mL}(\lambda_{\max}\|x\|_2^2+\sigma^2)^{mL}} e^{-\text{tr}\left(y(\lambda_{\min}x'x+\sigma^2 I_L)^{-1}y'\right)} e^{-\text{tr}\left(\bar{H}x(\lambda_{\min}x'x+\sigma^2 I_L)^{-1}x'\bar{H}'\right)} dP_o$$

$$\geq \int_{\|y\|_2 \leq \|x\|_2, \mathcal{R}(y) \subseteq \mathcal{R}(x)} \frac{k}{(\lambda_{\max}\|x\|_2^2+\sigma^2)^{mL}} dP_o,$$

where $k = k(\bar{H}, \Sigma)$ is a nonzero constant dependent on $\bar{H}$ and $\Sigma$. Assuming that

$$P_o(\|y\|_2 \leq \|x\|_2, \mathcal{R}(y) \subseteq \mathcal{R}(x)) = \Theta(e^{-l(y)}),$$

for some positive function $l : Y \to \mathbb{R}^+$, where $l(y) = \omega(\ln \|y\|_2)$, for sufficiently large $\|y\|_2^2$, we would obtain.

$$f_{P_o,Q}(y) \geq \Theta(e^{-2l(y)}).$$

But this means that

$$-\log_2 f_{P_o,Q}(y) = O(\min(\|y\|_2^2, l(y))).$$

Now, consider this together with (21), we can observe that for large values of $\|x\|_\eta^\eta$, the integral part of (21) is behaving as $O(\min(\|x\|_2^2, l(x)))$. Thus, a necessary condition for the capacity achieving-measure is $l(x) = \Omega(\|x\|_\eta^\eta)$. □

We remark that (21) in Proposition 5.3 provides necessary and sufficient condition for the capacity-achieving measure for Rician and Rayleigh channels (with full-rank covariance matrix) subject to any moment constraint of order $\eta > 1$. However, it is still not possible to solve (21) to find the capacity-achieving measure for the general choice of $\eta$, and this problem remains open for future investigations. For the case that $\eta = 2$, this problem has been addressed to some extent for Rayleigh channels in [15], [16], and [20]. For the case of SISO Rayleigh channels, it has been shown [16] that the capacity-achieving distribution has a finite number of mass points. Also, for the MIMO channel with isotropic Rayleigh distribution, the authors [15] have conjectured that the support of the capacity-achieving measure is in the form of concentric spheres around the origin with no interior point. More results can be found in [28].

# 6 Partial Channel State Information at the Receiver

In some fading channels, the system setup allows estimation of the channel at the receiver to some extent. In such cases, we assume that the channel state information is partially available at the receiver. Thus, in consideration of the general system model (1), we assume that the channel realization, $H$, is partially known at the receiver. We assume that the channel realization is governed by a Gaussian distribution characterized by $\bar{H}$ and $\Sigma$ (full-rank), as before. The channel state information is assumed to be available at the receiver in the form of elements from an LCH Borel-measurable space $V$, where each value $v \in V$ is an estimation for the channel realization $H$. We assume that $\boldsymbol{H} \times V$ is associated with a measure $Q \circ R$ which has a joint Gaussian density function. That is for every $v \in V$, the measure $Q_v$ has a Gaussian density function characterized by

$$\mu_{H|v} = \bar{H} + \Sigma_{Hv}\Sigma_{vv}^{-1}(v - \mu_v)$$

$$\Sigma_{H|v} = \Sigma - \Sigma_{Hv}\Sigma_{vv}^{-1}\Sigma_{vH},$$

where $\Sigma_{Hv}$ (and $\Sigma_{vH}$) denotes cross-covariance of $H$ and $v$, and $\mu_v$ and $\Sigma_{vv}$ are mean and covariance of the a Gaussian density function of the probability measure $R$ on $V$. To emphasize that the channel state information is partially known, we use a subscript "P", we denote the channel as $W_P(\mu_{H|v}, \Sigma_{H|v})$. To avoid extra difficulties, we assume that $\Sigma_{H|v}$ is full-rank for all $v \in V$.

Since the CSI is not fully known at the receiver, the size of block-length, $L$, is important in capacity analysis. Hence, we assume $X = \mathbb{C}^{n \times L}$. Note that $W_{Q_v}(\cdot|x) \ll T$ with a density function (obtained from (4))

$$f_{T,Q_v}(y|x) = \frac{\beta}{\det \Phi_{x,v}} \exp\left(\frac{\alpha^2 \|y\|_2^p - (y - \mu_{H|v}x)'\Phi_{x,v}^{-1}(y - \mu_{H|v}x)}{\sigma^2}\right) \tag{22}$$

where $\Phi_{x,v} = \frac{1}{\sigma^2}(x' \otimes I_m)\Sigma_{H|v}(x \otimes I_m) + I_{mL}$. For every input measure $P \in \mathscr{P}_{g,\Gamma}(X)$, we have $PW_{Q_v} \ll T$ with a density function

$$f_{T,P,Q_v}(y) = \int f_{T,Q_v}(y|x)dP. \tag{23}$$

As a result, the mutual information of this channel is expressed as

$$I(P, W_P(\mu_{H|v}, \Sigma_{H|v})|R) = \iiint f_{T,Q_v}(y|x) \log_2 f_{T,Q_v}(y|x) dT dP dR$$
$$- \iint f_{T,P,Q_v}(y) \log_2 f_{T,P,Q_v}(y) dT dR. \tag{24}$$

Note that the capacity per channel use is obtained by dividing the maximum of (24) by $L$.

## 6.1 Properties of mutual information

In Proposition 3.3 of Part I, we addressed the strict concavity of the mutual information in general. Since the strictness property is essential to addressing the uniqueness of the capacity-achieving probability measure, here, we address a simpler condition to verify this property of the mutual information function for fading channels with partial CSI at the receiver.

**Observation 6.1.** *For every $P \in \mathscr{P}_{g,\Gamma}(X)$, $f_{T,P,Q_v}(y)$ is continuous and nonzero for all $(y,v) \in Y \times V$.*

*Proof.* By (22), the continuity and positiveness of $f_{T,Q_v}(y|x)$ is obvious. The positiveness of $f_{T,Q_v}(y|x)$ implies the positiveness of $f_{T,P,Q_{\bar{H}}}(y)$. To prove the continuity, suppose that $(y_n, v_n) \to (y, v)$ in the Euclidean norm. Then,

$$\lim_n f_{T,P,Q_{v_n}}(y_n) = \lim_n \int f_{T,Q_{v_n}}(y_n) dP$$

$$= \lim_n e^{\frac{\alpha^2 \|y_n\|_2^p}{\sigma^2}} \lim_n \int \frac{\beta}{\det \Phi_{x,v_n}} e^{-\frac{(y_n - \mu_{H|v_n} x)' \Phi_{x,v_n}^{-1}(y_n - \mu_{H|v_n} x)}{\sigma^2}} dP$$

$$\text{By DCT } = e^{\frac{\alpha^2 \|y\|_2^p}{\sigma^2}} \int \frac{\beta}{\det \Phi_{x,v}} e^{-\frac{(y - \mu_{H|v} x)' \Phi_{x,v}^{-1}(y - \mu_{H|v} x)}{\sigma^2}} dP$$

$$= f_{T,P,Q_v}(y).$$

This proves the continuity of $f_{T,P,Q_v}(y)$ over $Y \times V$. □

We say two probability measures $P_1, P_2 \in \mathscr{P}_{g,\Gamma}(X)$ are *equivalent over* $W_P(\mu_{H|v}, \Sigma_{H|v})$ if they induce the same conditional output probability measure (conditional on $v$). Equivalently, two input measures are equivalent over $W_P(\mu_{H|v}, \Sigma_{H|v})$ if $f_{T,P_1,Q_v}(y) = f_{T,P_2,Q_v}(y)$ for all $(y,v) \in Y \times V$ such that $v$ is in the support of $R$.

**Proposition 6.1 (Strict concavity).** *The mutual information of channel $W_P(\mu_{H|v}, \Sigma_{H|v})$ is strictly concave with respect to the convex combination of two input measures $P_1$ and $P_2$, unless they are equivalent over $W_P(\mu_{H|v}, \Sigma_{H|v})$.*

*Proof.* By Observation 6.1, if there exists $y \in Y$ such that $f_{T,P_1,Q_v}(y) \neq f_{T,P_2,Q_v}(y)$ then there exists a neighborhood $U_{(y,v)} \subset Y \times V$ of $(y,v)$ with that property. Then by definition of $T \times R$ (3), $(T \times R)(U_{(y,v)}) > 0$. This implies that the set $U_{(y,v)}$ complies the requirements of Proposition 3.3 of Part I. Thus, the mutual information function of an PCSI channel is strictly concave with respect to the convex combination of $P_1$ and $P_2$ if and only if there exists $(y, v) \in Y \times V$ such that $f_{T,P_1,Q_v}(y) \neq f_{T,P_2,Q_v}(y)$. □

Another important property of mutual information which is useful in capacity analysis is its continuity, as shown below.

**Proposition 6.2 (Continuity).** *The mutual information of any channel $W_P(\mu_{H|v}, \Sigma_{H|v})$ is weak\* continuous over $\mathscr{P}_{g,\Gamma}(X)$.*

*Proof.* It suffices to check if the hypothesis of Theorem 3.3 of Part I is satisfied. We prove the theorem for $g(x) = \|x\|_2^2$, and the proof for other choices of $g$ is also similar. To prove hypothesis (a) of Theorem 3.3 of Part I, we proceed as the proof of Proposition 3.2. For every $P \in \mathscr{P}_{g,\Gamma}(X)$, let

$$A_c = \{(x, y, v) \in X \times Y \times V | f_{T,Q_v}(y|x) |\log_2 f_{T,Q_v}(y|x)| > c^2\}.$$

Let

$$B_c = \{(x, y, v) \in X \times Y \times V | f_{T,Q_v}(y|x) > c\},$$

for sufficiently large $c > 0$, it can be observed that $A_c \subset B_c$. Let

$$D_c = \{(x, y, v) \in X \times Y \times V | \alpha^2 \|y\|_2^p - \text{vec}\,(y - \mu_{H|v}x)' \Phi_{x,v}^{-1} \text{vec}\,(y - \mu_{H|v}x) > \sigma^2 \ln \frac{c \det \Phi_{x,v}}{\beta}\}.$$

It can be inspected that $B_c \subset D_c$. For sufficiently large $c > 0$, it can be inspected that $\alpha^2 \|y\|_2^p \geq \text{vec}\,(y - \mu_{H|v}x)' \Phi_x^{-1} \text{vec}\,(y - \mu_{H|v}x)$. Let also define

$$E_c = \{(x, y, v) \in X \times Y \times V | \|y\|_2^p > \frac{\sigma^2}{\alpha^2} \ln \frac{c \det \Phi_{x,v}}{\beta}\}.$$

Thus, for sufficiently large $c > 0$, $D_c \subseteq E_c$. Noting that $\det \Phi_{x,v} \geq 1$, let define $\omega(c) =$

$\left(\frac{\sigma^2}{\alpha^2}\ln\frac{c}{\beta}\right)^{1/p}$. In a similar discussion as in Proposition 3.2, we can deduce that

$$\iiint_{A_c} f_{T,Q_v}(y|x)\,|\log_2 f_{T,Q_v}(y|x)|\,dTdPdR$$

$$\leq \iiint_{E_c} f_{T,Q_v}(y|x)\,|\log_2 f_{T,Q_v}(y|x)|\,dTdPdR$$

$$\leq |\log_2 \beta|\iiint_{E_c}\frac{1}{\pi^{mL}\sigma^{2mL}\det\Phi_{x,v}}e^{-\frac{1}{\sigma^2}\text{vec}\,(y-\mu_{H|v}x)'\Phi_{x,v}^{-1}\text{vec}\,(y-\mu_{H|v}x)}d\mu_Y dPdR$$

$$+\frac{\log_2 e}{\sigma^2}\iiint_{E_c}\frac{2\alpha^2\|y\|_2^p}{\pi^{mL}\sigma^{2mL}\det\Phi_{x,v}}e^{-\frac{1}{\sigma^2}\text{vec}\,(y-\mu_{H|v}x)'\Phi_x^{-1}\text{vec}\,(y-\mu_{H|v}x)}d\mu_Y dPdR$$

$$\leq \left[\frac{|\log_2 \beta|}{\omega^2(c)}+\frac{2\alpha^2\log_2 e}{\sigma^2\omega^{2-p}(c)}\right]$$

$$\iiint\frac{\|y\|_2^2}{\pi^{mL}\sigma^{2mL}\det\Phi_{x,v}}e^{-\frac{1}{\sigma^2}\text{vec}\,(y-\mu_{H|v}x)'\Phi_{x,v}^{-1}\text{vec}\,(y-\mu_{H|v}x)}d\mu_Y dPdR$$

If we take $\sup_{P\in\mathscr{P}_{g,\Gamma}(X)}$ of both sides, the second term of the RHS is a finite value. Now, applying $\lim_{c\to\infty}$ to both sides, we observe that $\omega(c)\to\infty$ as $c\to\infty$. Hence,

$$\lim_{c\to\infty}\sup_{P\in\mathscr{P}_{g,\Gamma}(X)}\iiint_{A_c} f_{T,Q_v}(y|x)\,|\log_2 f_{T,Q_v}(y|x)|\,dTdPdR = 0.$$

This implies that the hypothesis (a) of Theorem 3.3 of Part I holds. Similar to the proof of Proposition 4.2, one can verify that the hypothesis (b) holds. Thus, both hypotheses of Theorem 3.3 of Part I hold, so the mutual information is weak* continuous. □

## 6.2 Capacity analysis

In this subsection, we address issues on the existence, the uniqueness, and the characterization of the capacity-achieving measures for fading channels with partial CSI at the receiver.

**Lemma 6.1 (Existence).** *For every channel $W_P(\mu_{H|v},\Sigma_{H|v})$, there exists a measure $P_o\in\mathscr{P}_{g,\Gamma}(X)$ that achieves the capacity of $W_P(\mu_{H|v},\Sigma_{H|v})$ over $\mathscr{P}_{g,\Gamma}(X)$.*

*Proof.* Proposition 5.2 states the weak* continuity of the mutual information over $\mathscr{P}_{g,\Gamma}(X)$. Since $\mathscr{P}_{g,\Gamma}(X)$ is weak* compact, by Proposition 4.1 of Part I, we conclude the assertion. □

Lemma 6.1 states the existence of the capacity-achieving measure over fading channels with partial CSI. One immediate result of strict concavity of mutual information, Proposition 6.1, is on the uniqueness of the capacity-achieving measure, as shown below.

**Lemma 6.2 (Uniqueness).** *For every channel $W_P(\mu_{H|v}, \Sigma_{H|v})$, the capacity-achieving measure over $\mathscr{P}_{g,\Gamma}(X)$ is unique up to the equivalency of input measures.*

*Proof.* The proof is essentially the same as the proof of Lemma 3.2. □

So far, in this section, we have shown the existence and the uniqueness of the capacity-achieving measure over fading channels with PCSI at the receiver. It remains to provide some insight to the characterization of such input measure.

**Proposition 6.3.** *Suppose $g(\cdot) = \|\cdot\|_\eta^\eta$ for $1 \leq \eta$. If $\eta > 2$, the capacity-achieving measure has a bounded support with no interior point. In contrast, if $1 \leq \eta \leq 2$, a necessary condition for the capacity-achieving measure is*

$$P_o(\|y\|_2 \leq \lambda_{\min}(v)\|x\|_2, \mathcal{R}(y) \subseteq \mathcal{R}(x)) = O(e^{-\alpha(v)\|y\|_\eta^\eta}),$$

*where $\mathcal{R}(\cdot)$ denotes the row space of a matrix, $\lambda_{\min}(v)$ denotes the minimum eigenvalue of $\Sigma_{H|v}$, and $\alpha : V \to \mathbb{R}^+$ is a positive function.*

*Proof.* Suppose $P_o W_{Q_v}$ is the optimal capacity-achieving output measure. Applying Kuhn-Tucker condition, Theorem 4.3 of Part I, to our problem, we need to find a positive value $\gamma > 0$ such that

$$\int D(W_{Q_v}(\cdot|x) \| P_o W_{Q_v}) dR - \gamma \|x\|_\eta^\eta \leq C - \gamma \Gamma.$$

We note that $W_{Q_v}(\cdot|x) \ll \mu_Y$ with density function

$$f_{Q_v}(y|x) = \frac{1}{\pi^{mL} \sigma^{2mL} \det \Phi_{x,v}} e^{-\frac{1}{\sigma^2} \text{vec}(y - \mu_{H|v} x)' \Phi_{x,v}^{-1} \text{vec}(y - \mu_{H|v} x)},$$

where $\Phi_{x,v}$ is defined as in (22). Using some straightforward mathematical manipulation, we obtain

$$-mL \log_2(\pi e \sigma^2) - \iint \log_2 \frac{d(P_o W_{Q_v})}{d\mu_Y} f_{Q_v}(y|x) d\mu_Y dR$$
$$- \int \log_2(\det \Phi_{x,v}) dR - \gamma \|x\|_\eta^\eta \leq C - \gamma \Gamma. \quad (25)$$

Now, the problem is to find an output density function together with the value $\gamma > 0$ such that the above inequality is satisfied with equality on all $x \in X$ in the support of the capacity-achieving measure. Unfortunately, because of the inherent difficulties in this expression, it is not possible to find an analytic solution for the output density function. However, through some asymptotic analysis discussion we can obtain some intuition on characterization of the support of the capacity-achieving measure, as follows.

*Case $\eta > 2$:* It can be inspected that for large values of $x$, the term $\gamma \|x\|_\eta^\eta$ is a dominant term. Hence, the first integral term in (25) must have a growth rate of $\|x\|_\eta^\eta$; otherwise, the support of the capacity-achieving measure is bounded. Thus, it suffices to study the asymptotic (tail) behavior of the density function $f_{P_o,Q_v}(y) = \frac{d(P_o W_{Q_v})}{d\mu_Y}$. For every fixed $y \in Y$, we have

$$f_{P_o,Q_v}(y) = \int \frac{1}{\pi^{mL} \sigma^{2mL} \det \Phi_{x,v}} e^{-\frac{1}{\sigma^2} \text{vec}(y - \mu_{H|v} x)' \Phi_{x,v}^{-1} \text{vec}(y - \mu_{H|v} x)} dP_o$$

$$\geq \int \frac{1}{\pi^{mL} \sigma^{2mL} \det \Phi_{x,v}} e^{-\frac{2}{\sigma^2} \text{vec} y' \Phi_{x,v}^{-1} \text{vec} y} e^{-\frac{2}{\sigma^2} \text{vec}(\mu_{H|v} x)' \Phi_x^{-1} \text{vec}(\mu_{H|v} x)} dP_o$$

$$\geq e^{-\frac{2}{\sigma^2} \|y\|_2^2} \int \frac{1}{\pi^{mL} \sigma^{2mL} \det \Phi_{x,v}} e^{-\frac{2}{\sigma^2} \text{vec}(\mu_{H|v} x)' \Phi_{x,v}^{-1} \text{vec}(\mu_{H|v} x)} dP_o$$

This means that $-\log_2 f_{P_o,Q_v}(y) = O(\|y\|_2^2)$. As a result, it can be verified that for $\eta > 2$, there exists no choice for the input measure so that the growth rate of the first integral of (25) catch up with the growth rate of $\|x\|_\eta^\eta$ for large values of $x$. Thus, the support of the capacity-achieving input measure is bounded for $\eta > 2$. Using an argument similar to the one in the proof of Proposition 5.3, one can deduce that the support of $P_o$ has no interior point.

*Case $\eta \leq 2$:* Recalling that $\Phi_{x,v} = \frac{1}{\sigma^2}(x' \otimes I_m) \Sigma_{H|v}(x \otimes I_m) + I_{mL}$, one can verify that for every $x \in X$,

$$\left(\frac{\lambda_{\min}(v)}{\sigma^2} x'x + I_L\right) \otimes I_m \leq \Phi_{x,v} \leq \left(\frac{\lambda_{\max}(v)}{\sigma^2} \|x\|_2^2 + 1\right) I_{mL},$$

where $\lambda_{\max}(v)$ and $\lambda_{\min}(v) > 0$ are the maximum and minimum eigenvalues of $\Sigma_{H|v}$. Let the operator $\mathcal{R}(\cdot)$ denote the row space of a matrix. Then,

$$f_{P_o,Q_v}(y) \geq \int \frac{1}{\pi^{mL} \sigma^{2mL} \det \Phi_{x,v}} e^{-\frac{2}{\sigma^2} \text{vec} y' \Phi_{x,v}^{-1} \text{vec} y} e^{-\frac{2}{\sigma^2} \text{vec}(\mu_{H|v} x)' \Phi_{x,v}^{-1} \text{vec}(\mu_{H|v} x)} dP_o$$

$$\geq \int \frac{e^{-\text{tr}(y(\lambda_{\min}(v) x'x + \sigma^2 I_L)^{-1} y')}}{\pi^{mL}(\lambda_{\max}(v) \|x\|_2^2 + \sigma^2)^{mL}} e^{-\text{tr}(\mu_{H|v} x(\lambda_{\min}(v) x'x + \sigma^2 I_L)^{-1} x' \mu'_{H|v})} dP_o$$

$$\geq \int_{\|y\|_2 \leq \lambda_{\min}(v) \|x\|_2, \mathcal{R}(y) \subseteq \mathcal{R}(x)} \frac{k(v)}{(\lambda_{\max}(v) \|x\|_2^2 + \sigma^2)^{mL}} dP_o,$$

where $k(v) = k(\mu_{H|v}, \Sigma_{H|v})$ is a nonzero function from $V$ to $\mathbb{R}^+$. Assuming that

$$P_o(\|y\|_2 \leq \lambda_{\min}(v) \|x\|_2, \mathcal{R}(y) \subseteq \mathcal{R}(x)) = \Theta(e^{-l(y)\alpha(v)}),$$

for some positive function $l : Y \to \mathbb{R}^+$, $\alpha : V \to \mathbb{R}^+$, where $l(y) = \omega(\ln \|y\|_2)$, we would obtain

$$f_{P_o,Q_v}(y) \geq \Theta(e^{-2l(y)r(v)}).$$

But this means that

$$-\log_2 f_{P_o,Q_v}(y) = O(\min(\|y\|_2^2, l(y)\alpha(v))).$$

Now, consider this together with (25), we can observe that for large values of $\|x\|_\eta^\eta$, the integral part of (25) is behaving as $O(\min(\|x\|_2^2, l(x)))$. Thus, a necessary condition for the capacity achieving-measure is $l(x) = \Omega(\|x\|_\eta^\eta)$. This concludes the assertion. □

We remark that (25) in Proposition 6.3 provides necessary and sufficient conditions for the capacity-achieving measure of Rician or Rayleigh channels (with full-rank covariance matrix) subject to any moment constraint $\eta > 1$. However, it is still not possible to solve (25) and determine the capacity-achieving measures. Hence, this problem remains open for future investigations.

# 7  Conclusion

In this part, we addressed a unified approach toward capacity analysis of multiple antenna channels. We used the results of the Part I to analyze the capacity of multiple antenna channels in a unified manner, irrespective of the type of fading, amount of correlation, and the amount of available knowledge about the channel state information (CSI) at the receiver. We studied the mutual information function and some of its analytical properties such as strict concavity and continuity for additive white Gaussian (AWGN) channels, fading channels with full CSI at the receiver, fading channels with no CSI, and fading channels with partial CSI at the receiver. Then, for each type of channels we studied the capacity value as well as issues such as the existence, uniqueness, and characterization of the capacity-achieving measures.

For channels with no CSI or partial CSI at the receiver, we provided necessary and sufficient conditions for characterization of the capacity-achieving measures and used asymptotic analysis to characterize the tail behavior of these measures. However, a closed form expres-sion or full characterization of these measures remain open for future investigations. As a direction for future research, one might consider the problem of characterization of the capacity-achieving measure for channels with no CSI or partial CSI at the receiver.